\documentclass[prd,preprint,eqsecnum,nofootinbib,amsmath,amssymb,tightenlines,dvips, superscriptaddress]{revtex4}
\preprint{JLAB-THY-11-1332}
\usepackage{bm, longtable}
\usepackage[pdftex]{graphicx}

\begin{document}
\baselineskip=20pt \title{Lattice QCD study of mixed systems of pions
  and kaons}

\author{William Detmold} \email{wdetmold@wm.edu}
\affiliation{Department of Physics, The College of William and Mary,
  Williamsburg, VA 23187-8795, USA.}  \affiliation{Jefferson
  Lab, 12000 Jefferson Avenue, Newport News, VA 23606, USA.}
\author{Brian Smigielski} \email{smigs@u.washington.edu}
\affiliation{Department of Physics, The College of William and Mary,
  Williamsburg, VA 23187-8795, USA.}  \affiliation{Department of
  Physics, National Taiwan University, Taipei, Taiwan.}

\begin{abstract}
  The different ground state energies of $N$-pion
  and $M$-kaon systems for $N+M\le12$ are studied in lattice
  QCD. These energies are then used to extract the various two- and
  three- body interactions that occur in these systems. 
  Particular attention is paid to additional thermal states 
  present in the spectrum because of the
  finite temporal extent. These calculations are 
  performed using one ensemble of 2+1 flavor
  anisotropic lattices with a spatial lattice spacing $a_s\sim
  0.125$~fm, an anisotropy factor $\xi=a_s/a_t=3.5$, and a spatial
  volume $L^3\sim (2.5\ {\rm fm})^3$.  The quark masses used correspond to pion
  and kaon masses of $m_{\pi} \sim 383 \ \mathrm{MeV}$ and $m_{K} \sim
  537 \ \mathrm{MeV}$, respectively. The isospin and strangeness 
  chemical potentials of these
  systems are found to be in the region where chiral perturbation
  theory and hadronic models predict a phase transition between a pion
  condensed phase and a kaon condensed phase.
\end{abstract}

\date\today
\maketitle

\section{Introduction}

An important goal of nuclear physics is to calculate nuclear
properties and processes from Quantum Chromodynamics (QCD).  The only
known way to perform such calculations in an {\it ab initio} manner is
using the non-perturbative formulation of QCD on a space-time lattice
(lattice QCD). Much progress has been made in recent years in studying
few hadrons systems, see for example the recent review of
Ref.~\cite{Beane:2010em}. Many channels of meson-meson, meson-baryon
and baryon-baryon interactions have been studied using the finite
volume behavior of two-particle energy levels
\cite{Hamber:1983vu,Luscher:1985dn,Luscher:1986pf,Luscher:1990ux}.
Ground state energies of three- \cite{Beane:2009gs} and four-
\cite{Yamazaki:2009ua} baryon systems have been computed for the first
time in QCD and quenched QCD, respectively.  To become a central part
of nuclear physics, the successes of lattice QCD in the realm of
few-hadron systems must be translated into the realm of many hadrons
and the complexity frontier must be addressed. To this end, recent
work has focused on numerical investigations in systems of up to twelve
pions or kaons
\cite{Beane:2007es,Detmold:2008fn,Detmold:2008yn,Detmold:2008bp}, and
algorithms \cite{Detmold:2010au} have been constructed to greatly
extend these studies. An important outcome of these studies was the
extraction of the two- and three- body interactions in pion and,
separately, kaon systems. The shifts in the $N$-meson energy levels at
finite volume were calculated numerically and then compared with the
analytical expectations derived in
Refs.~\cite{Beane:2007qr,Tan:2007bg,Detmold:2008gh}. These
calculations also probed the relation between isospin(hypercharge)
chemical potential and isospin(hypercharge) density, finding agreement
with the predictions of chiral perturbation theory
\cite{Son:2000xc,Kogut:2001id}.

In this work, we study more complicated multi-meson systems involving
both pions and kaons. As in the single species case, the expected
volume dependence of the energy of an $N$-$\pi^+$, $M$-$K^+$ system
has been computed \cite{Smigielski:2008pa} and these calculations will
enable us to extract the scattering lengths for the three two-body
interactions, $\bar{a}_{\pi^+ \pi^+}$, $\bar{a}_{\pi^+ K^+}$,
$\bar{a}_{K^+ K^+}$, and also to study the various zero momentum
three-body interactions, parameterized as $\bar{\eta}_{3,\pi \pi
  \pi}$, $\bar{\eta}_{3,\pi \pi K}$, $\bar{\eta}_{3,\pi KK}$,
$\bar{\eta}_{3,KKK}$ (restricted to the maximal isospin in each case).
Multi-meson systems such as these exhibit a rich phase structure
\cite{Dashen:1974ff,Son:2000xc} and are phenomenologically relevant in
a number of settings ranging from RHIC \cite{Klein:2003fy} to the
interiors of neutron stars \cite{Kaplan:1986yq}. At low temperatures,
the ground-state of a finite density system of, for example, pions is
expected to undergo a transition from a Bose-Einstein condensed phase
\cite{Sawyer:1972cq,Scalapino:1972fu,Baym:1973zk,Dashen:1974ff,Campbell:1974qt,Campbell:1974qu}
to a BCS-type superconducting phase as the isospin chemical potential
is increased, and other interesting phenomena such as
$\rho$-condensation may also occur
\cite{Voskresensky:1997ub,Sannino:2002wp,Aharony:2007uu}. Similar
effects are expected in kaon systems. The phase structure of a mixed
system of pions and kaons is a less well investigated
question. Studies in the NJL (Nambu--Jona-Lasinio) model,
\cite{Barducci:2004nc,He:2005nk,Warringa:2005jh}, random matrix theory
\cite{Arai:2009zzb}, and SU(3) chiral perturbation theory ($\chi$PT)
with non-zero isospin and hypercharge chemical potentials
\cite{Kogut:2001id}, suggest that pion and kaon condensed phases
compete in interesting ways for different ranges of isospin and
hypercharge chemical potentials.

In the following, we focus on numerical analysis of lattice QCD
realizations of these complex multi-meson systems and on the
methodology needed to extract the two- and three-body interactions.
We calculate the 90 possible two-point correlation functions involving
$N$-$\pi^+$s and $M$-$K^+$s for all $N+M\le12$. These results are then
analyzed to extract the ground state energies of the appropriate
quantum numbers, taking care to account for thermal excitations in the
lattice volume. Finally, the resulting large set of energies is used
to extract the seven underlying interactions discussed above. The
extracted energies also allow us to study the chemical potential of
these systems. This study is performed using a single ensemble of 2+1
flavor anisotropic gauge field configurations generated by the Hadron
Spectrum Collaboration \cite{lin-2009-79} and so should be viewed as
an exploratory study, testing methods necessary for analysis of
complicated many body systems. Future work with different lattice
spacings, volumes and quark masses will allow contact with experiment
\cite{Pislak:2001bf,Pislak:2003sv,Adeva:2005pg,Batley:2005ax,Adeva:2000vb},
in the case of two-body interactions, and with other lattice studies.

\section{Multi-hadron correlation functions}
In order to extract the energies of the multi meson systems, we will
study the behavior of two point correlation functions
\begin{equation}
  C_{N,M}(t)=\Big\langle 0 \Big| \Big(\sum_{\bm{x}} \pi^{-}(\bm{x},t) \Big)^N
  \Big( \sum_{\bm{x}} K^{-}(\bm{x},t) \Big)^M  \Big(\pi^{+}(\bm{0},t)
  \Big)^N \Big(K^{+}(\bm{0},t) \Big)^M  \Big| 0 \Big\rangle\,,
  \label{eq:Cnm}
\end{equation}
where the pion and kaon interpolating operators are defined in terms
of quark fields as $\pi^{-\dagger}=\pi^+({\bf x},t)=\overline{u}({\bf
  x},t)\gamma_5d({\bf x},t)$ and $K^{-\dagger}=K^+({\bf
  x},t)=\overline{u}({\bf x},t)\gamma_5d({\bf x},t)$ , respectively.
This system corresponds to a system of $N \pi^{+}$'s and $M K^{+}$'s
with total momentum, $\bm{P}_{\mathrm{tot}}=0$.

It is instructive to first consider the form of such
correlation functions which can be generically expressed as
\begin{equation}
  C(t)=\langle 0 | \hat{\mathcal{O}}^{\prime\dagger}(t) \hat{\mathcal{O}}(0) |
  0 \rangle\,,
\end{equation}
for some interpolating operators, ${\cal O}^{(\prime)}$, with
commensurate quantum numbers.\footnote{The creation and annihilation
  interpolating operators are not in general Hermitian conjugates of
  one another.}  In a system with infinite (Euclidean) temporal
extent, application of the transfer matrix formalism shows that
\begin{equation}
  C(t) = \sum_{k=0}^{\infty} Z_k e^{-E_k t}\,,
\end{equation}
where the sum runs over all eigenstates with the quantum numbers of
the operators under study.  The $E_k$ are the energies of eigenstates
$|E_k \rangle$ and $Z_k= \langle 0 |
\hat{\mathcal{O}}^{\prime\dagger}(0) | E_k \rangle \langle E_k |
\hat{\mathcal{O}}(0) | 0 \rangle$.  At large Euclidean time, the
correlator is dominated by the ground state energy of the system,
$C(t) \to Z_{0} \ \exp({-E_0} t)$.  However, in a system with a finite
temporal extent, $T$, a natural choice of temporal boundary conditions
for quark and gluon fields, anti-periodic and periodic respectively,
will force $C(t)$ to be periodic (assuming the operators under study
are bosonic, as in Eq.~(\ref{eq:Cnm})). The expected form of the
correlation function can again be inferred:
\begin{align}\label{corrFxn}
  C(t)&=\frac{1}{\mathcal{Z}} \mathrm{tr}\left(e^{-T\hat{H}}
    \hat{\mathcal{O}}^{\prime\dagger}(t) \hat{\mathcal{O}}(0) \right)
  = \frac{1}{\mathcal{Z}}\sum_{n,m=0}^{\infty} e^{-(T-t)E_n}e^{-t
    E_m}\langle E_n | \hat{\mathcal{O}}^{\prime\dagger}(0) | E_m
  \rangle \langle E_m | \hat{\mathcal{O}}(0) | E_n \rangle \,,
\end{align}
with $\mathcal{Z}=\mathrm{tr}(e^{-T \hat H})$. The contributions on
the right-hand side of Eq.~(\ref{corrFxn}) can be separated into two
distinct classes. The first class is defined by choosing $|E_n\rangle$
to be the vacuum state ($E_n=0$) and correspond to the ground state
and excited states of the system. These contributions are special as
they persist in the $T \to \infty$ (or zero temperature) limit. The
remaining contributions exist solely due to the finite time extent 
and are known as thermal states \cite{Beane:2007es,Detmold:2008yn,
Beane:2009kya, Richards:2000hp,Gockeler:2001db,Prelovsek:2008rf,
Detmold:2008ww}. These states vanish as $T \to \infty$
because of the first exponential under the sum in Eq.~(\ref{corrFxn}).

Thermal contributions can be illustrated with a concrete
example. Consider $C_{4,0}(t)$ which corresponds to a system 
with the quantum numbers of four pions. 
The correlation function is dominated by a sum of three
contributions in the large $t$ limit:
\begin{align}\label{eq:C40}
  C_{4\pi,0K}(t)&=Z_1 e^{-E_{4\pi}T/2} \ \cosh\left( E_{4\pi} t_T \right) + Z_2 e^{-(E_{3\pi}+M_{\pi})T/2}\ \cosh\left( (E_{3\pi}-M_{\pi}) t_T \right) \nonumber \\
  &+ Z_3\cosh\left(E_{2\pi}T \right) + \cdots \nonumber \\
  &=\tilde{Z_1} \ \cosh\left( E_{4\pi} t_T \right) + \tilde{Z_2} \
  \cosh\left( (E_{3\pi}-M_{\pi}) t_T\right)+ \tilde{Z_3} + \cdots \,,
\end{align}
where $t_T = t-T/2$, the ellipsis denotes contributions involving
excited pion systems that we will ignore in the current discussion,
and the $\tilde{Z}$'s are constant with respect to time. The first
term in Eq.~(\ref{eq:C40}) corresponds to states where all four
$\pi^{-}$'s propagate forward in time ($|E_n\rangle=|0\rangle,\,
|E_m\rangle=|4\pi\rangle$) and four $\pi^{+}$'s propagate backward in
time ($|E_n\rangle=|4\pi\rangle,\, |E_m\rangle=|0\rangle$). The second
term represents three $\pi^{-}$'s propagating forward in time and one
$\pi^{+}$ propagating backward in time ($|E_n\rangle=|\pi\rangle,\,
|E_m\rangle=|3\pi\rangle$) as well as three $\pi^{+}$'s propagating
backward in time and one $\pi^{+}$ propagating forward in time
($|E_n\rangle=|3\pi\rangle,\, |E_m\rangle=|\pi\rangle$). Finally, the
last term arises from two $\pi^{-}$'s propagating forward in time and
two $\pi^{+}$'s propagating backward in time
($|E_n\rangle=|2\pi\rangle,\, |E_m\rangle=|2\pi\rangle$).

The general form for the correlation function of $N$-pions and
$M$-kaons can be straightforwardly worked out and is given by the
following
\begin{align}\label{eq:Cnmhadronic}
  C_{N,M}(t) &= \sum_{m=0}^M \sum_{n=0}^N Z^{N-n,M-m}_{n,m}
  \cosh\left( \Delta E^{N-n,M-m}_{n,m} t_T \right) +
  Z^{\frac{N}{2},\frac{M}{2}}_{\frac{N}{2},\frac{M}{2}} \delta_{N
    \bmod{2},0} \delta_{M \bmod{2},0} + \cdots \,,
\end{align}
where $\Delta E^{N-n,M-m}_{n,m} = (E_{N-n,M-m}-E_{n,m})$, the ellipsis
denotes excited state contributions, and the last term in
Eq.~(\ref{eq:Cnmhadronic}) is only present when $N,M$ are even.  It is
apparent that the number of possible terms contributing to
$C_{N,M}(t)$ grows with $N$ and $M$. Locality of the transfer matrix
guarantees that the eigen-energies $E_{n,m}$ appearing in many places
in different $C_{N,M}(t)$ are identical. Consequently, multiple
correlation functions (choices of $N$ and $M$) can be used to extract
the common set of eigen-energies.

Thermal states contribute to most lattice calculations and need to be
considered in precise analyses. They are particularly prevalent in the
multi-hadron context as these systems easily factorize into multiple
color singlet states that can propagate over long distances. In
Refs.~\cite{Beane:2007es,Detmold:2008fn,Detmold:2008yn}, the issue of
thermal states was avoided by using Dirichlet boundary conditions. In
principle, such boundary conditions introduce unknown contamination
into correlation functions, however these works analyzed correlations
significantly separated from the boundary to reduce such effects.

\section{Details of Lattice Calculation}
\label{sec:lattice}

The lattice calculations presented in this work are based on a single
ensemble of anisotropic gauge field configurations generated by the
Hadron Spectrum Collaboration~\cite{lin-2009-79}. These configurations
were generated using a stout smeared \cite{PhysRevD.69.054501} gauge
action and a 2+1 flavor clover fermion action; further details can be
found in Refs.~\cite{lin-2009-79}. The spatial and temporal lattice
spacings are $a_s=0.123 \ \mathrm{fm}$ and $a_t=a_s/\xi$ respectively,
where $\xi=3.5$ is the renormalized anisotropy. The lattices have a
volume of $20^3\times128$, corresponding in physical units to $(2.5\
{\rm fm})^3\times 4.6\ {\rm fm}$, and the light and strange quark
masses are such that the pion and kaon have masses of $m_{\pi}\sim 383
\ \mathrm{MeV}$ and $m_K\sim 537 \ \mathrm{MeV}$.  These
configurations have been extensively studied by the HSC and NPLQCD
collaborations. For this study, an ensemble of 400 configurations were
chosen from a long stream of $\sim$10,000 trajectories and are well
separated in Monte-Carlo time such that autocorrelations are reduced.

Our analysis makes use of light and strange quark propagators
generated by the NPLQCD collaboration. For each configuration,
randomly positioned APE \cite{PhysRevD.56.2743,Gusken:1989ad} smeared
sources were used to generate approximately 75 propagators. The EigCG
inverter \cite{Stathopoulos:2007zi} was used to perform the multiple
inversions efficiently.  These propagators were then APE smeared at
the sink and combined into the correlation functions of
Eq.~(\ref{eq:Cnm}).

Naively, the number of contractions involved in the correlation
functions of Eq.~(\ref{eq:Cnm}) is enormous, $(N+M)!N!M!$.  To compute
the requisite contractions, we extend the methods developed in
Ref.~\cite{Detmold:2008fn} to the mixed species case.  Following the
manipulations of Ref.~\cite{Detmold:2008fn} it is straightforward to
show that for arbitrary $12 \times 12$ matrices, $\Pi, K$:
\begin{align}\label{det1}
  \det\Big(1+ \lambda_P \Pi + \lambda_K K \Big)&= \exp\Big(
  \mathrm{tr} \Big[ \sum_{j=1}^{\infty} \frac{(-)^{j-1}}{j} \left(
    \lambda_P \Pi +
    \lambda_K K  \right)^j \Big] \Big)\nonumber \\
  &=\exp\Big( \mathrm{tr} \Big[ \sum_{j=1}^{\infty}
  \frac{(-)^{j-1}}{j} \sum_{k=1}^{j} \frac{j!}{k!(j-k)!}  \lambda^k_P
  \lambda^{j-k}_K \Pi^k
  K^{j-k} \Big] \Big)\nonumber \\
  &=1+\lambda_P \ \mathrm{tr} \ \Pi + \lambda_K \ \mathrm{tr} \ K +
  \frac{\lambda^2_P}{2} \Big( \Big[ \mathrm{tr} \ \Pi \Big]^2 -
  \mathrm{tr} \ \Pi^2  \Big) \nonumber \\
  &+ \frac{\lambda^2_K}{2} \Big( \Big[ \mathrm{tr} \ K \Big]^2 -
  \mathrm{tr} \ K^2 \Big) + \lambda_P \lambda_K \Big( \mathrm{tr} \
  \Pi
  \ \mathrm{tr} \ K - \mathrm{tr} \ \Pi K \Big) \nonumber \\
  & + \cdots,
\end{align}
with
\begin{eqnarray}
  \Pi(t) &=& \sum_{\bm{x}} S_u(\bm{x},t;\bm{0},0) \
  S_d^{\dagger}(\bm{x},t;\bm{0},0), \\
  K(t) &=& \sum_{\bm{x}} S_u(\bm{x},t;\bm{0},0) \ S_s^{\dagger}(\bm{x},t;\bm{0},0),
\end{eqnarray}
where the $S_f$ correspond to the source and sink smeared quark
propagators of flavor $f$. Additionally, one can generalize the
results of the single species case to show that in the mixed species
case,
\begin{equation}\label{det2}
  \det\Big(1+ \lambda_P \Pi + \lambda_K K  \Big) = \frac{1}{12!} \sum_{j=1}^{12} \sum_{k=0}^j \binom{12}{j}
  \binom{j}{k}\lambda_P^k \lambda_K^{j-k} C_{k,j-k}(t)\,,
\end{equation}
where
\begin{equation}
  C_{k,j-k}(t) = \epsilon^{\alpha_1 \dots \alpha_k \mu_1 \dots \mu_{j-k} \xi_1 \dots \xi_{12-j}}
  \epsilon_{\beta_1 \dots \beta_k \nu_1 \dots \nu_{j-k} \xi_1 \dots \xi_{12-j}} \Pi_{\alpha_1}^{\beta_1}\dots
  \Pi_{\alpha_k}^{\beta_k} K_{\mu_1}^{\nu_1} \dots K_{\mu_{j-k}}^{\nu_{j-k}}.
\end{equation}
By expanding the exponential to a particular order in $\lambda_{P}$
and $\lambda_K$, and equating Eq.~(\ref{det1}) and Eq.~(\ref{det2}),
one can identify the function $C_{N,M}(t)$.

The computational complexity of these two-species contractions is
significantly more than that of the single species cases studied
previously. The reader can find an example, the $C_{4\pi,3K}(t)$
correlator, in Eq. (\ref{multiMesonCorrFxnEx}). As in the single
species cases \cite{Detmold:2008fn,Detmold:2008yn}, high precision
arithmetic is required to perform these contractions correctly and
this is implemented using the {\tt arprec} and {\tt qd} libraries
\cite{Bailey:2002ab}.  Explicit calculations for all $N+M=13$
correlation functions show these correlators vanish to the requisite
precision and are an effective check of the correctness of our code.
The computational cost of computing all contractions $C_{N,M}(t)$ for
$N+M\le12$ and all $0\le t<128$ is approximately twenty minutes on a
single core, compared to a few seconds for the single species case.
While not available at the time that the current calculations began,
the recursive constructions developed in Ref.~\cite{Detmold:2010au}
provide an alternate, and computationally more efficient, way of
generating the contractions. The results of both methods agree to
arbitrary precision.\footnote{In future calculations that extend the
number of pions and kaons beyond twelve, these recursive techniques
will be critical as the explicit code generated for all the
requisite contractions would be unmanageably large.}
  
\section{Volume Dependence of Multi-hadron energies}

As discussed above, the calculations of the correlators in
Eq.~(\ref{eq:Cnmhadronic}) determine the energies of the mesonic
systems. These can in turn be used to determine the interactions
through the well-known results of L\"uscher
\cite{Luscher:1985dn,Luscher:1986pf} in the two body case and the
results of
Refs.~\cite{Beane:2007qr,Tan:2007bg,Detmold:2008gh,Smigielski:2008pa}
in the many meson case where a perturbative expansion in the inverse
volume is performed.

\subsection{Volume dependence of two particle energies}
\label{sec:volume-depend-multi-nonpert}

An extraction of the scattering lengths from the single species and
mixed species two particle systems provides a baseline reference with
which to compare the results of the multi-particle analysis. In the
center-of-mass frame, we define the interaction momentum,
$p=|\bm{p}|$, from the energy shift $\Delta E_{AB} \equiv
E_{AB}-E_A-E_B= \sqrt{p^2+m_A^2}+\sqrt{p^2+m_B^2}-m_A-m_B$.  $\Delta
E_{AB}$ is determined from lattice calculations and one first solves
for the momentum $p$ \cite{Luscher:1986pf}.  
L\"uscher's formula relates this momentum to
the phase-shift, $\delta(p)$, of two particle scattering as
\begin{align}\label{eq:nonpert-Lusc}
  p \cot \delta(p) &= \bm{S}\Big( \frac{p^2 L^2}{4\pi^2} \Big)\,,
\end{align}
where $\bm{j} \in \mathbb{Z}^3$ and
\begin{align}
  \bm{S}(x) &= \sum_{\bm{j}}^{|\bm{j}|<\Lambda_k}
  \frac{1}{|\bm{j}|^2-x}-4\pi \Lambda_k,
\end{align}
is a regulated three dimensional zeta function. By expanding the
left-hand side of Eq.~(\ref{eq:nonpert-Lusc}) using the
effective-range expansion, $p \cot \delta(p) = -\frac{1}{a} +
\frac{r_0}{2}p^2$, the scattering length, $a$, can be determined. In
the absence of interactions, $\bm{S}$ possesses poles at
$\bm{j}=2\pi\bm{n}/L$ for $\bm{n} \in \mathbb{Z}^3$. With interactions
these poles are shifted.  One then computes $a$ as:
\begin{equation}\label{npScatt}
  a= -\frac{\pi L}{\bm{S}\Big( \frac{p^2 L^2}{4\pi^2} \Big)}\,.
\end{equation}
Using this formula with the associated one-body and two-body energies
will yield precise determinations of the scattering lengths as the
one- and two- body energies are the most cleanly determined and the
above results only omit exponentially suppressed finite volume
effects.

\subsection{Volume dependence of multi-meson energies}
\label{sec:volume-depend-multi}

In Ref.~\cite{Smigielski:2008pa}, the energy shift of a system
$N$-pions and $M$-kaons in a finite volume from the corresponding
non-interacting system was calculated. The shift is given by:
\begin{align}\label{singleSpeciesAnswer}
  \Delta{E}(n,m,L)&={E}(n,m,L)-n \ m_{\pi}-m \
  m_{K}\nonumber\\
  &=\Delta E_{\pi}(n,L)+\Delta E_{K}(m,L)+\Delta \tilde{E}_{\pi
    K}(n,m,L)\,,
\end{align}
with $m_{\pi{K}}=m_{\pi}m_{K}/(m_{\pi}+m_{K})$ and $i\in{\pi,K}$ and
\begin{align}\label{singleSpeciesShift}
  \Delta E_{i}(n,L)&=\frac{4\pi{\bar{a}_{ii}}}{m_{i}L^3}
  \left(\begin{array}{c}n\\2\end{array}\right)\left[1
    -\left(\frac{\bar{a}_{ii}}{\pi{L}}\right)\mathcal{I}
    +\left(\frac{\bar{a}_{ii}}{\pi{L}}\right)^{2}\left(\mathcal{I}^2
      +(2n-5)\mathcal{J}\right)\right
  .\nonumber\\
  &\left.-\left(\frac{\bar{a}_{ii}}{\pi{L}}\right)^3\left(\mathcal{I}^3+(2n-7)
      \mathcal{I}\mathcal{J}+(5n^2-41n+63)\mathcal{K}\right)\right]\nonumber\\
  &+\left(\begin{array}{c}n\\3\end{array}\right)\frac{\bar{\eta}_{3,iii}(\mu)}{L^6}
  +\mathcal{O}(L^{-7}) \,,
\end{align}
\begin{align}\label{piKanswer}
  \Delta \tilde{E}_{\pi
    K}(n,m,L)&=\frac{2{\pi}\bar{a}_{\pi{K}}mn}{m_{\pi{K}}L^3}
  \left[1-\left(\frac{\bar{a}_{\pi{K}}}{\pi{L}}\right)\mathcal{I}\right.\nonumber\\
  &\left.+\left(\frac{\bar{a}_{\pi{K}}}{\pi{L}}\right)^{2}
    \left(\mathcal{I}^2+\mathcal{J}\left[-1+2\frac{\bar{a}_{\pi\pi}}{\bar{a}_{\pi{K}}}(n-1)
        \left(1+\frac{m_{\pi
              K}}{m_{\pi}}\right)\right.\right.\right.\nonumber\\
  &+2\left.\left.\left.\frac{\bar{a}_{KK}}{\bar{a}_{\pi{K}}}(m-1)\left(1+\frac{m_{\pi
              K}}{m_{K}}\right)\right]\right)\right.\nonumber\\
  &\left.+\left(\frac{\bar{a}_{\pi{K}}}{\pi{L}}\right)^{3}\left(-\mathcal{I}^{3}
      +f^{\mathcal{K},\pi{K}}\left(\frac{\bar{a}_{\pi\pi}\bar{a}_{KK}}{\bar{a}_{\pi{K}}^2}
      \right)\mathcal{K}\right.\right.\nonumber\\
  &+\left.\left.\sum_{i=0}^{2}\sum_{p=\pi,K}\left(f^{\mathcal{I}\mathcal{J},p}_{i}
        \mathcal{I}\mathcal{J}+f^{\mathcal{K},p}_{i}\mathcal{K}\right)\left(
        \frac{\bar{a}_{pp}}{\bar{a}_{\pi{K}}}\right)^{i}\right)\right]\nonumber\\
  &+\frac{nm(n-1)\bar{\eta}_{3,\pi\pi{K}}(L)}{2L^6}+\frac{nm(m-1)
    \bar{\eta}_{3,\pi{KK}}(L)}{2L^6}+\mathcal{O}(L^{-7}),
\end{align}
where the parameters $\bar{a}_{ij}$ are related to the scattering
lengths and effective ranges through
\cite{Detmold:2008fn,Smigielski:2008pa}:
\begin{align}\label{scattLengths}
  a_{\pi\pi}=\bar{a}_{\pi\pi}-\frac{2{\pi}\bar{a}_{\pi\pi}^{3}r_{\pi\pi}}{L^3},\qquad
  a_{KK}=\bar{a}_{KK}-\frac{2{\pi}\bar{a}_{KK}^{3}r_{KK}}{L^3},\qquad
  a_{\pi{K}}=\bar{a}_{\pi{K}}-\frac{2{\pi}\bar{a}_{\pi{K}}^{3}r_{\pi{K}}}{L^3}.
\end{align}
It is the $\bar{a}_{ij}$ parameters that will be determined in the
current lattice calculations.  The four volume dependent (but
renormalization scale independent) quantities characterizing the
momentum independent three-body interactions are defined by
($y=m_{\pi}/m_{K}$):
\begin{align}\label{3bodyForces}
  \bar{\eta}_{3,iii}(L)&=\eta_{3,iii}(\mu)+\frac{64\pi{a_i}^4}{m_i}(3\sqrt{3}-4\pi){\rm
    log}(\mu{L})-\frac{96a_i^4}{{\pi^2}m_i}(2q[1,1]+r[1,1]), \nonumber\\
  \bar{\eta}_{3,\pi\pi{K}}(L,y)&=\eta_{3,\pi\pi{K}}(\mu,y)
  -\frac{4a_{\pi{K}}^4}{{\pi^2}m_{\pi{K}}}\sum_{i=0}^{2}\sum_{p=\pi,K}\sum_{\mathcal{N}\in\mathcal{N}_1}
  \left(\frac{a_{p}}{a_{\pi{K}}}\right)^{i}f_{i}^{\mathcal{N},p}\mathcal{N},\nonumber\\
  \bar{\eta}_{3,\pi{KK}}(L,y)&=\eta_{3,\pi{KK}}(\mu,y)
  -\frac{4a_{\pi{K}}^4}{{\pi^2}m_{\pi{K}}}\sum_{i=0}^{2}\sum_{p=\pi,K}\sum_{\mathcal{N}\in\mathcal{N}_2}
  \left(\frac{a_{p}}{a_{\pi{K}}}\right)^{i}f_{i}^{\mathcal{N},p}\mathcal{N},
\end{align}
and
\begin{align}
  \mathcal{N}_1&=\left\{\hat{Q}(1,y),\hat{Q}(y,1),\hat{R}(y,1),\hat{R}(1/y,1/y)\right\},\nonumber\\
  \mathcal{N}_2&=\left\{\hat{Q}(1,1/y),\hat{Q}(y,y),\hat{R}(y,y),\hat{R}(1,1/y)\right\},
\end{align}
with $i=\pi,K$.  The functions $\hat{Q}$, $\hat{R}$, $q$, and $r$
along with the coefficients $f_i$ are defined in
Ref.~\cite{Smigielski:2008pa}. The finite parts of $\hat{Q}(a,b)$ and
$\hat{R}(a,b)$ are scheme dependent quantities where changes in the
value will be compensated by changes in $\eta_3(\mu)$; the numerical
values for the Minimal Subtraction (MS) scheme are given in
Ref.~\cite{Smigielski:2008pa}.  However, the $\bar{\eta}_3$ are
scheme independent and these are the quantities that will be determined
during a lattice calculation.  Furthermore, the three-body
interactions in the $\pi\pi{K}$ and $\pi{KK}$ cases depend on the mass
ratio, $m_{\pi}/m_K$.  Finally, in the limit of $N \to 0$ or $M\to0$,
this result simplifies to the previously determined $N$-boson case
while the limit $M \to N$ with $m_{K}\to m_{\pi}$ and all interactions
set to be equal it simplifies to the $2N$-boson case.

\section{Analysis Strategies}
\label{fitStrategies}

After calculating the above correlation functions on the ensemble of
gauge configurations, we obtain lattice measurements of $C_{N,M}(t)$
in Eq.~(\ref{eq:Cnm}). Provided that we can reliably deal with the
presence of thermal and excited states, these allow us to extract the
ground state energies of the $N$-pion, $M$-kaon systems. By matching
these measurements onto the analytic expectations of
Eq.~(\ref{piKanswer}), the various two- and three-body interaction
parameters can then be extracted. There are many ways in which such an
analysis could proceed. One could attempt to directly fit the
correlation functions in terms of overlap factors, and the scattering
parameters and three-body interactions by inserting the explicit form
of the energies from Eq.~(\ref{piKanswer}). However, we choose to
perform this as a two stage analysis, first extracting the various
energies by fits to the correlation functions using the model
functions of Eq.~(\ref{eq:Cnmhadronic}) and then subsequently fitting
these energies to the analytic forms of Eq.~(\ref{piKanswer}) to
extract the interaction parameters. The primary advantage of this
approach to performing the fits is increased stability, however, care
must be taken to preserve the significant correlations between the
different correlation functions which are computed on the same set of
gauge configurations.

\subsection{Wavefunction Overlap}
\label{sec:wavefxnOverlap}

As shown in Eq.~(\ref{eq:Cnmhadronic}), the $Z$-factors for each term
are in general distinct. This makes for a difficult analysis since
there are large numbers of linear and non-linear fit parameters that
must be determined. In the non-interacting system, these
$Z$-factors are simply related. In the single-species case, the
relation would be $Z^{N-n}_{n}=\binom{N}{n}Z_N$ such that $Z_N$ is
common to each term. The factorial simply counts the number of ways
that $n$-pions propagate forward and $(N-n)$ propagate
backward. Generalizing this to the multi-meson case leads to
$Z^{N,M}_{n,m}=\binom{N}{n}\binom{M}{m}Z_{N,M}$. In an interacting
system, however, this relationship does not hold.  Nevertheless,
in this work, we study pion and kaon systems with relatively weak
interactions, and therefore we employ the ansatz above for the
wavefunction factors.  We stress this is a crucial assumption,
allowing us to subsequently fit all 90 correlators. As a check,
analyses were also performed allowing the $Z$-factors to remain
distinct for as many correlators as possible. Results from both
methods were found to agree within uncertainty, but a full analysis
using independent $Z_{n,m}^{N-n,M-m}$'s proved unfeasible.

An alternative method that can be utilized for these systems is the
method of Variable Projection (VARPRO)
\cite{Kaufman:1975,Fleming:2004hs} which eliminates linear
fit parameters by performing their minimization analytically.  This approach
is useful in the case of unrelated $Z^{N-n,M-m}_{n,m}$'s from
Eq.~(\ref{corrFxn}).  Additionally, when many energies exist to be
fitted, convergence to the minimum of parameter space is not always
guaranteed, particularly if the minimum is flat. To aid convergence,
Bayesian priors \cite{Lepage:2001ym,Howson:2005} can be implemented so
the physically reasonable region of parameter space is immediately
tested. In an earlier stage of our analysis, the VARPRO 
method (with priors) was used for correlators with lower 
total particle number.  The results were found to agree 
within uncertainty with the results using the ansatz above.

As a final check, the fitted energies were substituted back into the
form for the correlator and compared against the original data set. 
Consistency was achieved within uncertainty. 
Therefore, for the remainder of this work, we assume the 
simplified form for the wavefunction overlap.

\subsection{Thermal effects of Correlation Functions}
\label{sec:thermal-effects}

A particular multi-meson correlator depends not only on its ground
state energy in the large-$T$ limit, but also upon fewer body energies
in the thermal states that were discussed earlier.  To account for
this, we began by fitting the one pion(kaon) correlator to determine
$E_{1,0}$($E_{0,1}$) with a bootstrap method as will be explained
below. Once these values were known, we turned to the two pion(kaon)
correlator. This correlator depends on $E_{2,0}$($E_{0,2}$) and
$E_{1,0}$($E_{0,1}$). Instead of refitting both energies, we used
previous bootstrapped values of $E_{1,0}$($E_{0,1}$) in the two
pion(kaon) fit function. This is reasonable since the cleanest signal
for the $E_{N,M}$ energy will come from its respective correlator. To
preserve all covariances in the data, the same bootstrap sample of
gauge field configurations was used for all correlators. This
sequential procedure was used to build up the analysis and ascertain
all of the multi-meson energies. The fitted results for the energies
and their statistical and systematic errors can be found in Tables
\ref{singleSpeciesEnergyTable} and \ref{multiSpeciesEnergyTable}.

\subsection{Statistical Analysis}
\label{sec:stat-anal}

The correlation functions span the time interval $[0,127]$ and a
suitable subinterval, where the fit formula is applicable, must be
chosen.  As a first step, the (periodic) correlation function data was
reflected around its midpoint in order to improve
statistics. Effective mass plots for all correlation functions were
constructed, allowing initial estimates to be made of where (in time)
contamination from excited states ceases.  These plots revealed that
excited state contributions are suppressed by $t \sim 20$. Therefore,
in an effort to be as general as possible, fits were run on all time
intervals, $(t_{\mathrm{min}},t_{\mathrm{min}} + \Delta t)$, such that
$t_{\mathrm{min}} \in [25,33]$ and $\Delta t \in [10,20]$.  Using
intervals beyond these values showed no improvement in fit results.

In this work, we utilized the bootstrap method of statistical
resampling to estimate uncertainties. To construct the bootstrap
ensembles, we first averaged over correlators from all sources on each
gauge configurations, for each $N,M$.  Let $C_k (t_i)$ be a given
correlation function for some fixed number of pions and kaons, $N$ and
$M$, calculated on a given gauge field configuration $k$ at time
$t_i$. Then we denote this set as $\lbrace C_{1}(t_i), C_2(t_i),
\cdots, C_{\mathcal{M}}(t_i) \rbrace$, with $\mathcal{M}=400$. The
data covariance matrix is defined from the full data set as:
\begin{equation}\label{fullDataCovMatrix}
  \mathcal{C}(t_i,t_j) = \frac{1}{\mathcal{M}
    \left(\mathcal{M}-1\right)}\sum_{k=1}^{\mathcal{M}}\left(C_{k}(t_i)
    - \bar{C}(t_i) \right) \left(C_{k}(t_j) - \bar{C}(t_j) \right),
\end{equation} 
where $\bar{C}(t_i)=(1/\mathcal{M})\sum_{k=1}^{\mathcal{M}} C_k
(t_i)$.

Next, we randomly sampled configurations from the full data set and
constructed a bootstrap sample, $B_1(t_i)$ of $\mathcal{M}$ elements,
for each time $t_i$. This sample was averaged to yield
$\bar{B}_1(t_i)$ and this entire process repeated $\mathcal{P}$ times
where $\mathcal{P}=450$, resulting in: $\lbrace \bar{B}_1(t_i), \bar{B}_2(t_i), \cdots ,
\bar{B}_{\mathcal{P}}(t_i)\rbrace$. For each of the $ \bar{B}_j$, a
$\chi_j^2$ function was determined as:
\begin{equation}
  \chi^2_j = \sum_{t,t'} \Big( \bar{B}_j(t) - y(t;E_{N,M}) \Big) \ \mathcal{C}^{-1}(t,t') \ \Big( \bar{B}_j(t') - y(t';E_{N,M}) \Big),
\end{equation}
where $y(t;E_{N,M})$ represents the fit model in
Eq.~(\ref{eq:Cnmhadronic}).  Each of the $\chi^2_j$ functions were
minimized and results recorded.

In order to assess the reliability of the fits, we define the goodness-of-fit, $Q_j =
\frac{1}{2^{d/2}\Gamma(d/2)} \int_{(\chi^2_{\mathrm{min}})_j}^{\infty}
d\chi_j^2 \ (\chi_j^2)^{d/2-1}e^{-\chi_j^2 / 2}$, 
where $d$ is the number of degrees-of-freedom, 
for each $\chi_j^2$. Once a best fit energy is selected
on each bootstrap sample, we determine a weighted 
average for the ensemble,
\begin{align}
  \langle E_{N,M} \rangle &=
  \frac{1}{\mathcal{W}_1}\sum_{j=1}^{\mathcal{P}} Q_j E_{N,M}^{(j)}
  ,\end{align} with weights $Q_j$, and a weighted standard deviation,
\begin{align}
  \sigma^2_{N,M} &= \frac{\mathcal{W}_1^2}{\mathcal{W}_1^2 -
    \mathcal{W}_2} \ \sum_{j=1}^{\mathcal{P}}Q_j \Big( E_{N,M}^{(j)} -
  \langle E_{N,M} \rangle \Big)^2\,.
\end{align}
Above, $\mathcal{W}_1 = \sum_{j=1}^{\mathcal{P}} Q_j$ and
$\mathcal{W}_2 = \sum_{j=1}^{\mathcal{P}} Q_j^2$.

To assign a systematic uncertainty for the extracted energies, we
return to the fits performed to the correlators. This is a necessary
component of the analysis since there exists arbitrariness in the
choice of which $\mathrm{t_{min}}$ and which $\Delta t$ were selected
for the correlation function fits as explained above. The systematic
error on the best fit energy is defined as
\begin{equation}
  \sigma_{\mathrm{sys}}=\frac{1}{2}\Big| E_{N,M}(\Delta t_{+}) - E_{N,M}(\Delta t_{-})  \Big|,
\end{equation}
where $E_{N,M}(\Delta t_{+,-})$ are the energies extracted from
$C_{N,M}(t)$ using the time intervals
$[\mathrm{t_{min}}+1,\mathrm{t_{max}}+1]$ and
$[\mathrm{t_{min}}-1,\mathrm{t_{max}}-1]$.

\subsection{Two and Three Body Parameters}
\label{sec:2-3bodyParams}

The second stage of our analysis now focuses on the extraction of the
scattering lengths and three-body coefficients from the measured
energies using Eqs.~(\ref{singleSpeciesShift}) and (\ref{piKanswer}).
These determinations rely on a second bootstrap analysis involving a
resampling of the extracted energies. The bootstrapping procedure
for a specific correlation function yielded $\mathcal{P}$ energies,
and these formed the bootstrap samples for the extraction of
the two and three-body parameters.

Once the best fit multi-meson energies were known, a very similar
procedure used for the analysis of the correlators was used to find
the $\bar{a}$'s and $\bar{\eta}$'s. Since a bootstrap ensemble exists
for every best fit energy value, we created an energy sample,
$\mathcal{\bm{E}}_{\alpha}$ such that $\alpha \in
[1,\mathcal{P}]$. This sample carries an additional vector index that
labels the energies within the vector. In the case of single
species pion energies (the kaon case is identical), an energy vector
initially composed of $\mathcal{\bm{E}}_{\alpha}=\lbrace
E^{(\alpha)}_{2,0},E^{(\alpha)}_{3,0},E^{(\alpha)}_{4,0} \rbrace$ was
used to fit to $\bar{a}_{\pi\pi}$ and $\bar{\eta}_{3,\pi\pi\pi}$.  We
included another energy and refitted the interaction parameters and
repeated this until all the energies were exhausted.  In the
multi-species case, a base set of $\lbrace
E^{(\alpha)}_{2,0},\cdots,E^{(\alpha)}_{12,0}, E^{(\alpha)}_{0,2},
\cdots, E^{(\alpha)}_{0,12} \rbrace$ along with ten randomly selected
multi-species energies was created and fits performed for all seven
hadronic parameters. This first set thus made use of 34 different
energies. This set was enlarged by one, the parameters were
refitted, and the process repeated until all ninety energies were
used. The energy covariance matrix used in these fits, is defined
according to:
\begin{equation}\label{energyCovMatrix}
  \mathcal{C}(\mathcal{E})_{i,j}=\frac{1}{\mathcal{P}-1}\sum_{\alpha=1}^{\mathcal{P}}\left( \mathcal{E}_{\alpha,i} - \langle \mathcal{E}_{i} \rangle  \right) \left( \mathcal{E}_{\alpha,j} - \langle \mathcal{E}_{j} \rangle \right),
\end{equation}
such that $\langle \mathcal{E}_{i} \rangle =
(1/\mathcal{P})\sum_{\alpha=1}^{\mathcal{P}}\mathcal{E}_{\alpha,i}$,
and the energy $\chi^2$ on each bootstrap is defined as:
\begin{equation}
  \chi^2_{\alpha} = \sum_{i,j} \Big( \mathcal{E}_{\alpha,i} - f_{\alpha,i}(\bar{a},\bar{\eta}) \Big) \ \mathcal{C}(\mathcal{E})^{-1}_{i,j} \ \Big( \mathcal{E}_{\alpha,j} - f_{\alpha,j}(\bar{a},\bar{\eta}) \Big),
\end{equation}
where $f(\bar{a},\bar{\eta})$ is shorthand notation for the fit
functions in Eqs.~(\ref{singleSpeciesAnswer})-(\ref{piKanswer}).

The systematic errors assigned to the $\bar{a}$'s and $\bar{\eta}$'s
are more complicated than those of the energies.  Given a particular
energy set of $\mathcal{N}$ energies that are used to make a
determination of $\bar{a}$'s and $\bar{\eta}$'s there are
$3^{\mathcal{N}}$ different combinations of the intervals that must be
fit in order to completely propagate the systematic uncertainties of the
energies to those of the interaction parameters (it is $3^\mathcal{N}$
because there is a $[\mathrm{t_{min}},\mathrm{t_{max}}]$ for each best
fit energy as well as its systematic counterparts corresponding to the
shifted time interval in the forward and backward direction).  Even in the
single species case, when $\mathcal{N}=10$, there are already $\sim 6
\times 10^4$ combinations.  For the multi-species case, it is too
costly to fit all these permutations. Rather, we only fit
$\mathcal{O}(10^3)$ randomly chosen permutations and take the
difference of the mean of this set from the best fit $\bar{a}$ and
$\bar{\eta}$ as the systematic error.  From fitting all permutations
in the single species case, up to $\mathcal{N}=9$, it was seen the
systematic error stabilized well before the total number of
combinations was computed and we assume this is also the case for the
two-species case.

\section{Results}

\subsection{Energies}
\label{res:energies}

\begin{figure}[htb]
  \centering
  \includegraphics[scale=0.70,clip=true]{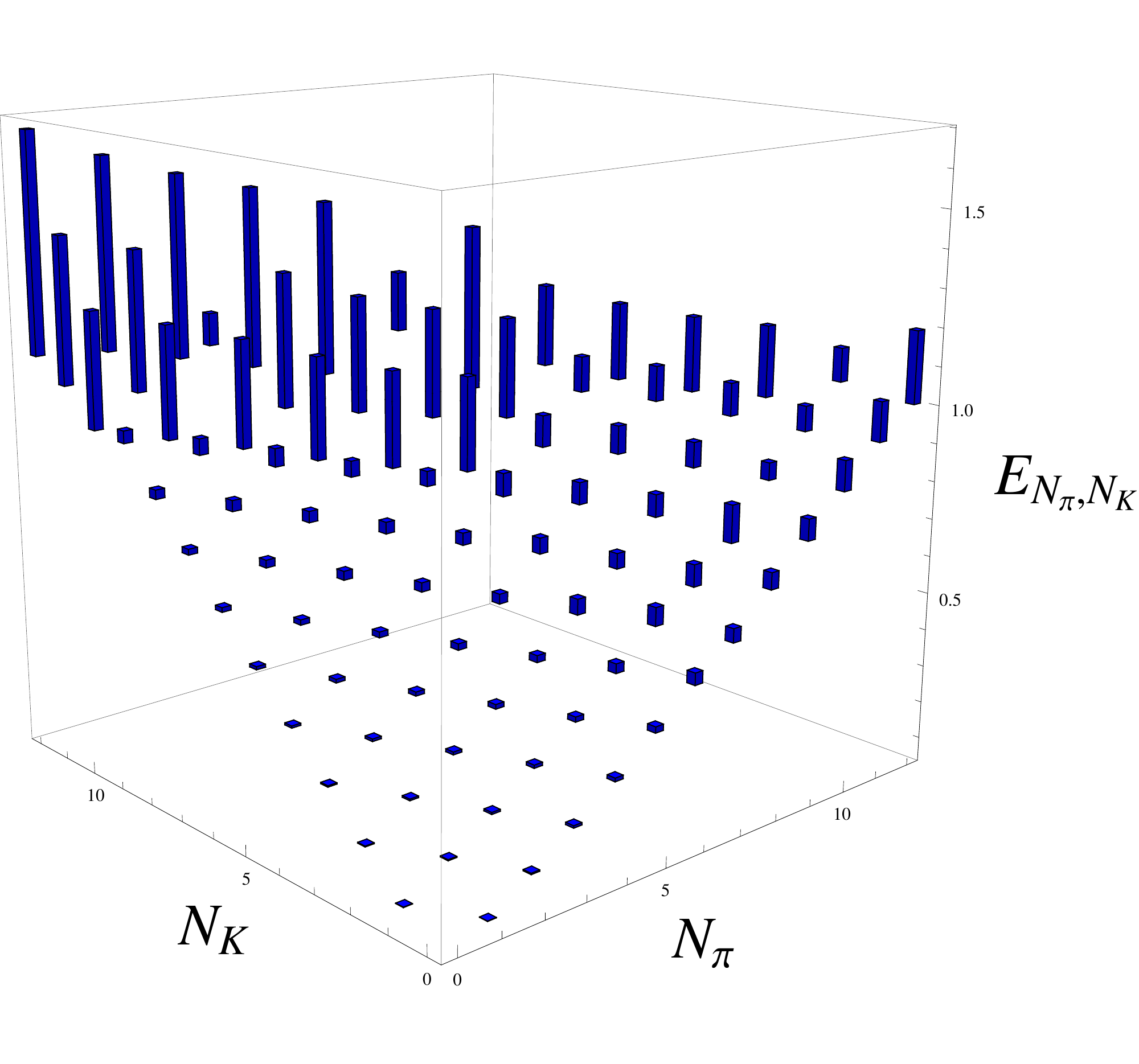}
  \caption{Energy of multi meson states. Uncertainties shown are result from
    combining statistical and systematic uncertainties in quadrature.}
  \label{energyPlot}
\end{figure}

Using the methods discussed above, we extracted the energies of
the mixed and pure species system, from all ninety correlators.  The
final extracted values are shown in Tables
\ref{singleSpeciesEnergyTable} and \ref{multiSpeciesEnergyTable}
below, along with their associated fit ranges. These energies
are shown in a three-dimensional plot along with their
respective uncertainties in Fig.~\ref{energyPlot}. 

\begin{figure}[h]
\begin{center}$
\label{energies:correlatorPlots}
\begin{array}{cc}
\includegraphics[scale=0.42]{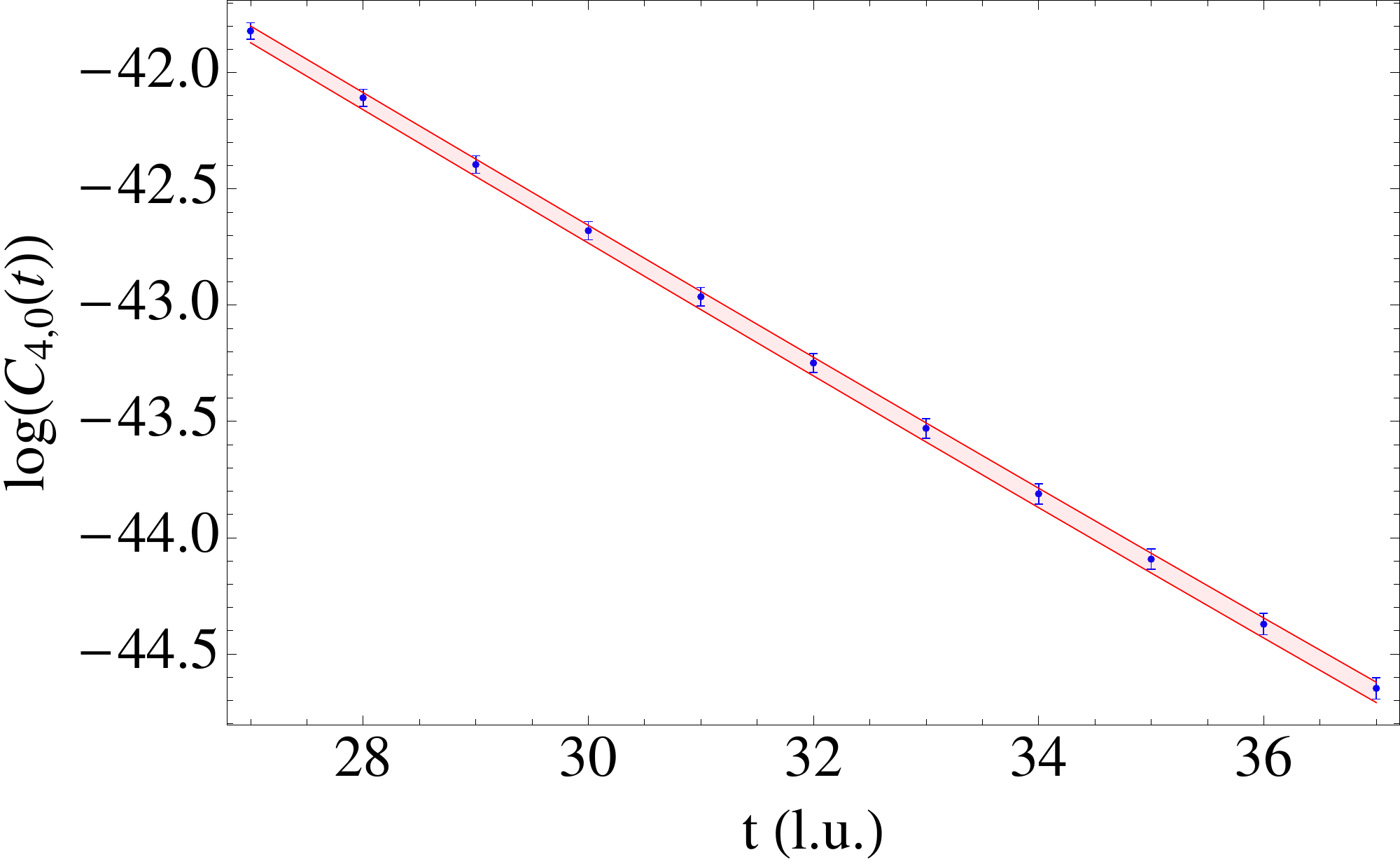} &
\includegraphics[scale=0.42]{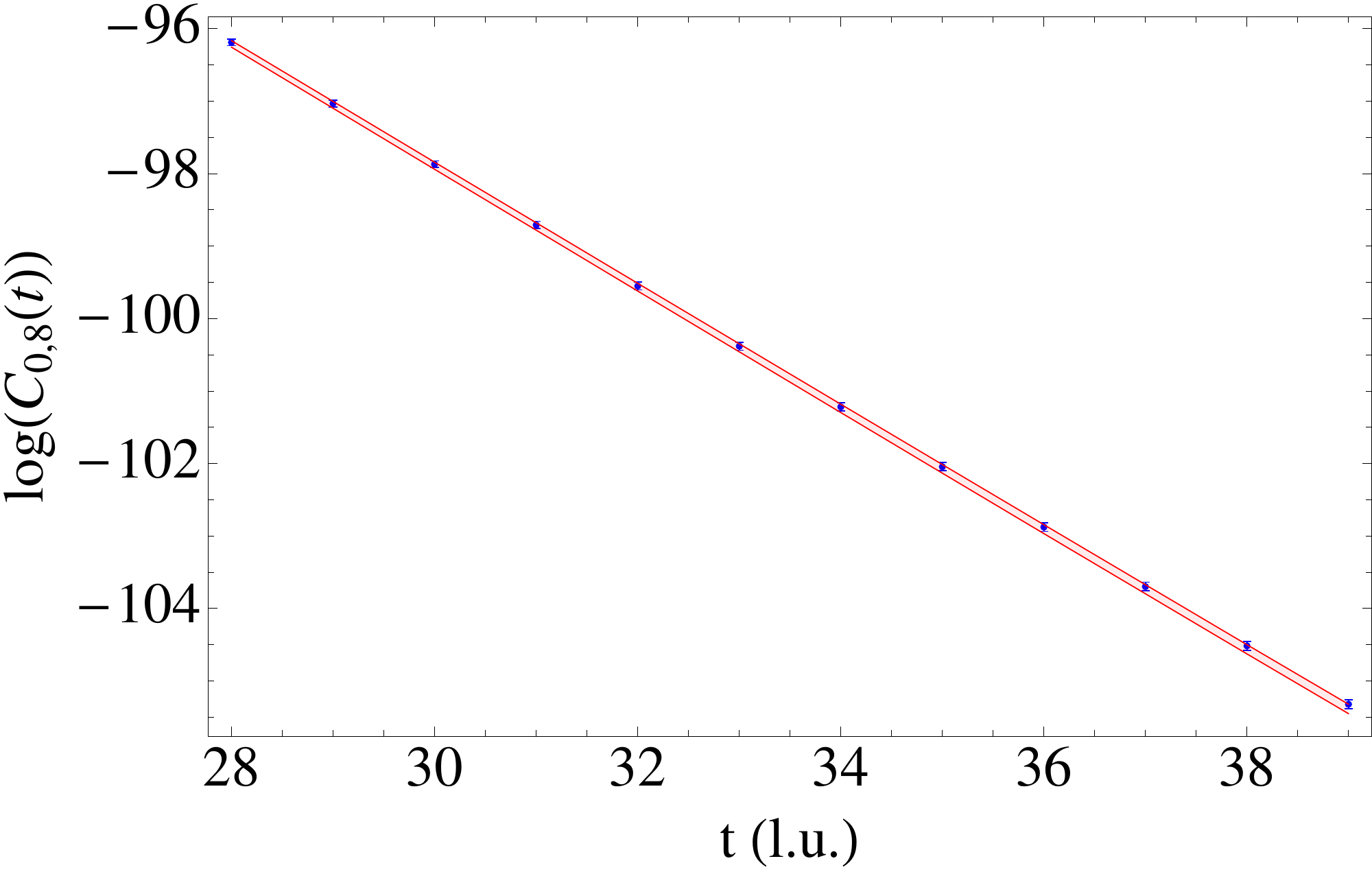} \\ 
\includegraphics[scale=0.42]{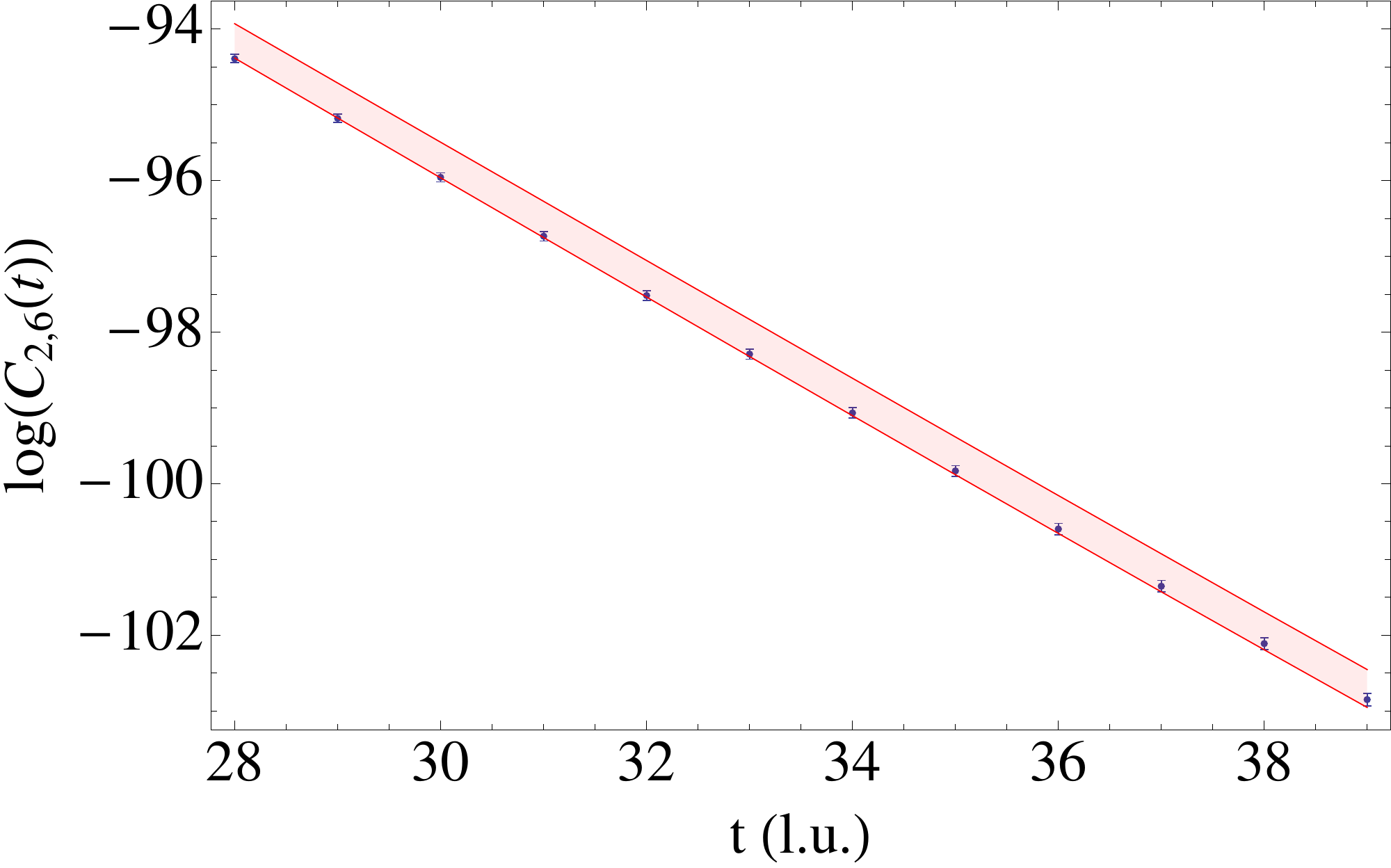} &
\includegraphics[scale=0.42]{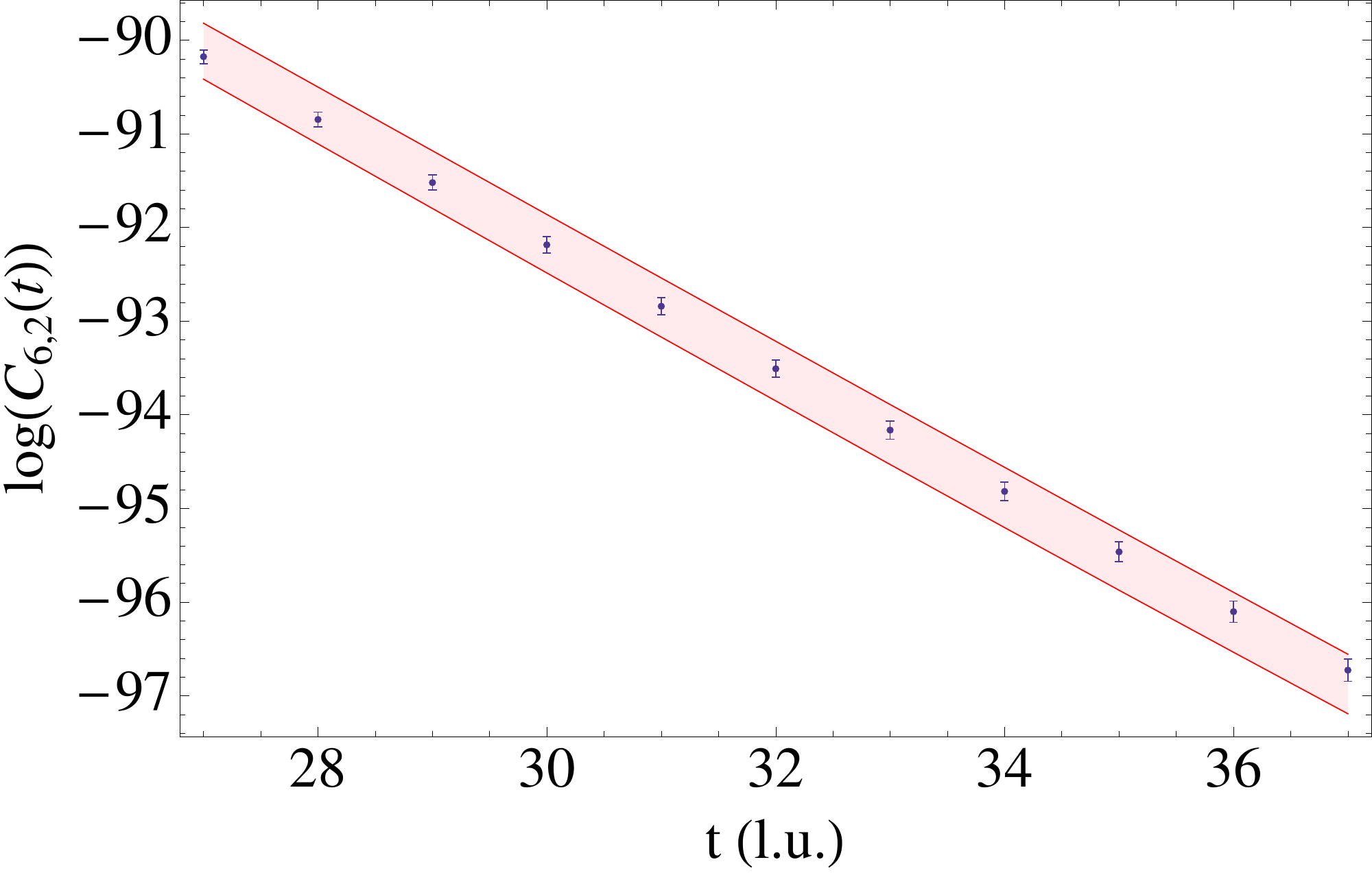} 
\end{array}$
\end{center}
\caption{Plots of the log of the fitted correlation function (red) and those
based on the full data set of gauge configurations (blue) for their respective
fit intervals for representative $N$ and $M$. The red
envelope denotes the uncertainty in the fitted correlator propagated 
from the uncertainty of the energies.}
\end{figure}

The fits become progressively more difficult as the 
number of mesons grows because of
the increasing thermal contamination. This is directly reflected in
the quality of the fits decreasing for large meson number in both
the pure species and mixed species case. Fits to example
correlators are shown in Fig.~\ref{energies:correlatorPlots}.

\subsection{Interactions}
\label{res:interactions}

The extractions of interaction parameters from mixed meson energies
were performed to yield the three scattering lengths and four
three-body coefficients. This work builds upon the studies of
\cite{Beane:2007es,Detmold:2008fn, Detmold:2008yn} and presents the
first measurements of $\bar{\eta}_{3,\pi\pi K}$, and $\bar{\eta}_{\pi KK}$
since these parameters can only be measured within the framework of
the mixed-meson system.

\begin{figure}[htb]
  \centering
  \includegraphics[scale=0.60,clip=true]{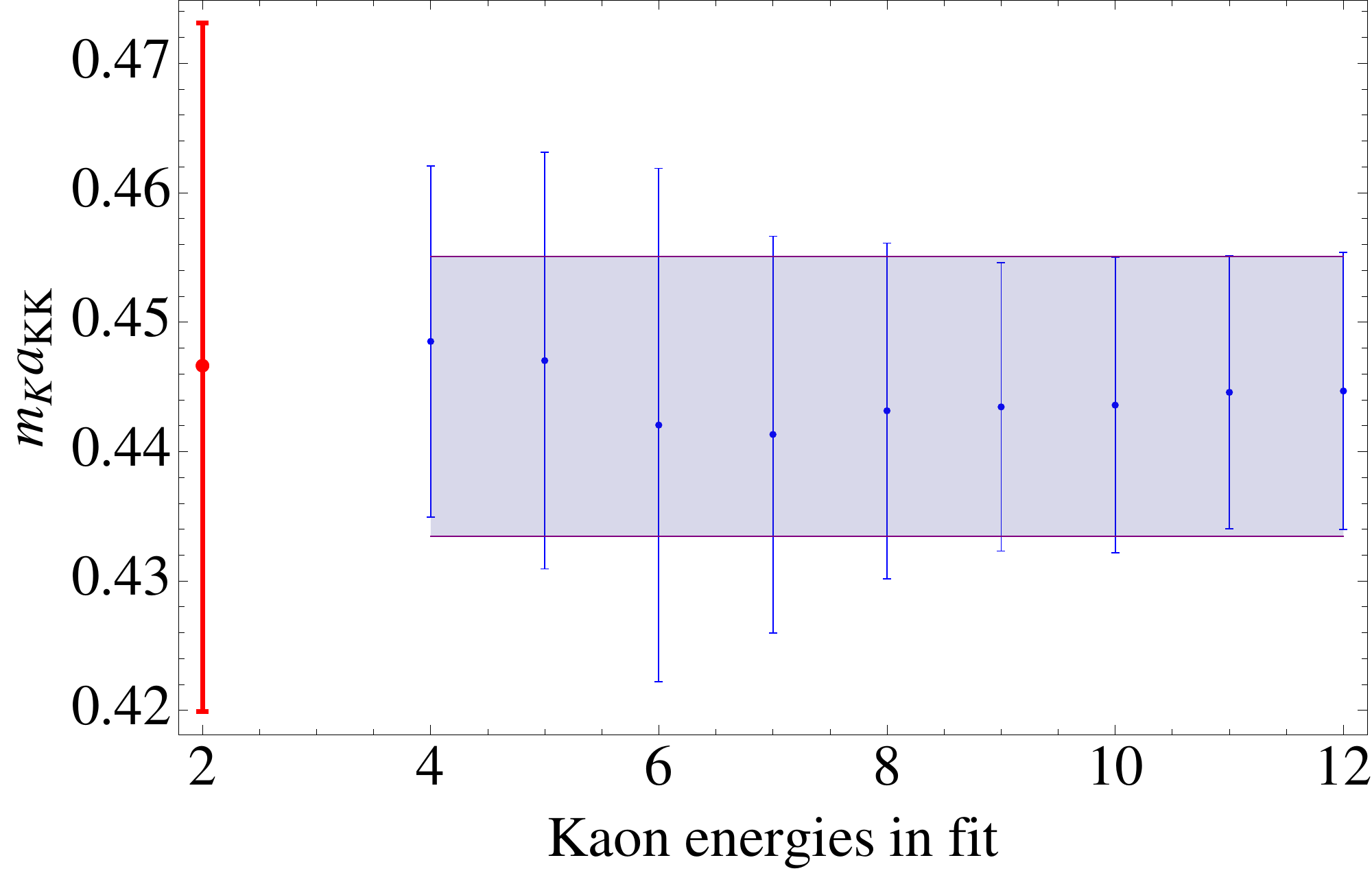} \\
  \includegraphics[scale=0.60,clip=true]{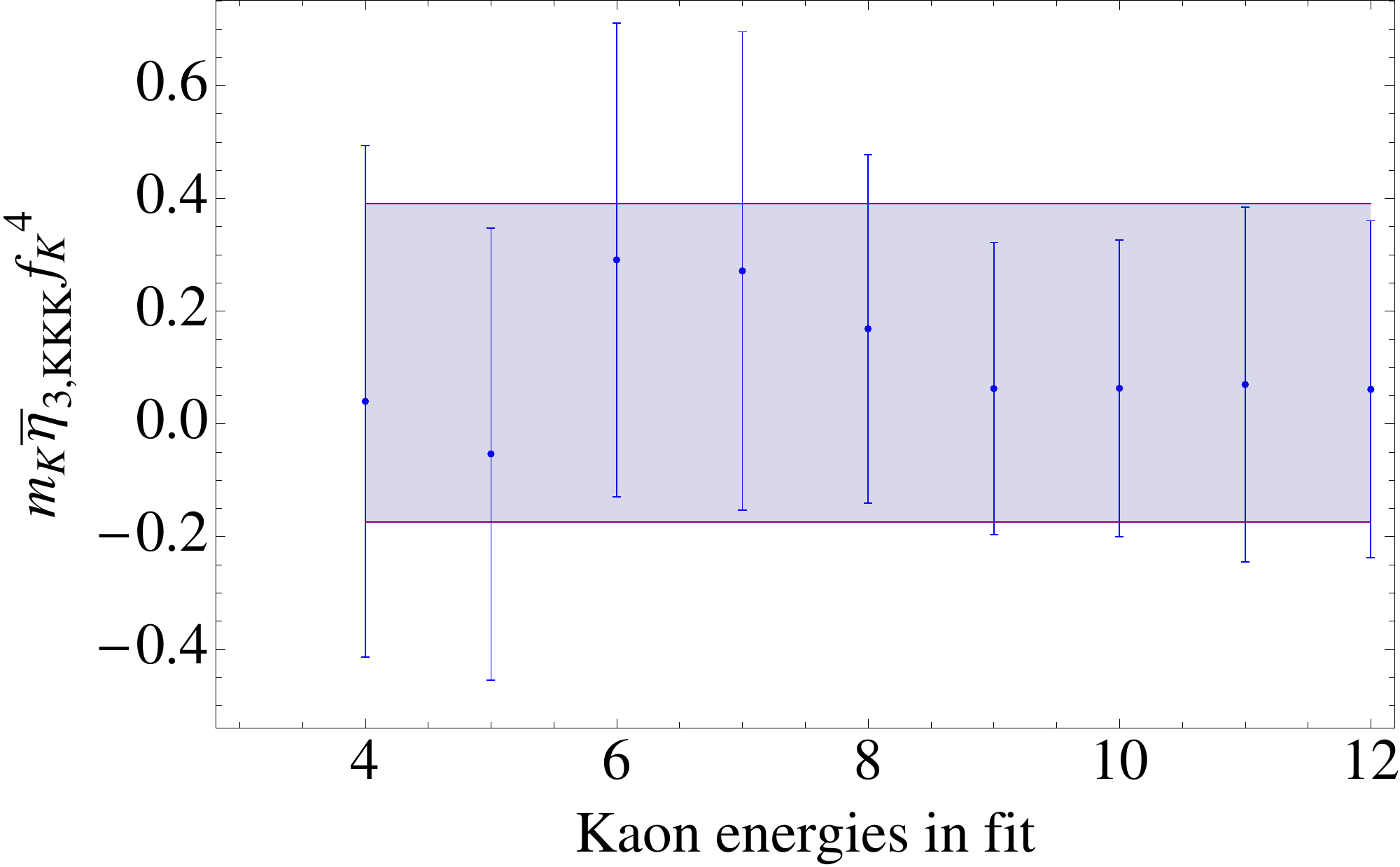}
  \caption{Calculation of scattering lengths and three-body
    coefficients for kaons from Eq.~(\ref{singleSpeciesAnswer}).
    Uncertainties per extraction result from combining statistical 
    and systematic uncertainties in quadrature.
    The shaded band defines the standard deviation of the 
    mean of all extractions. The red point is the nonperturbative
    L\"uscher result.}
  \label{singleMesonScattParams2}
\end{figure}

\begin{figure}[htb]
  \centering
  \includegraphics[scale=0.60,clip=true]{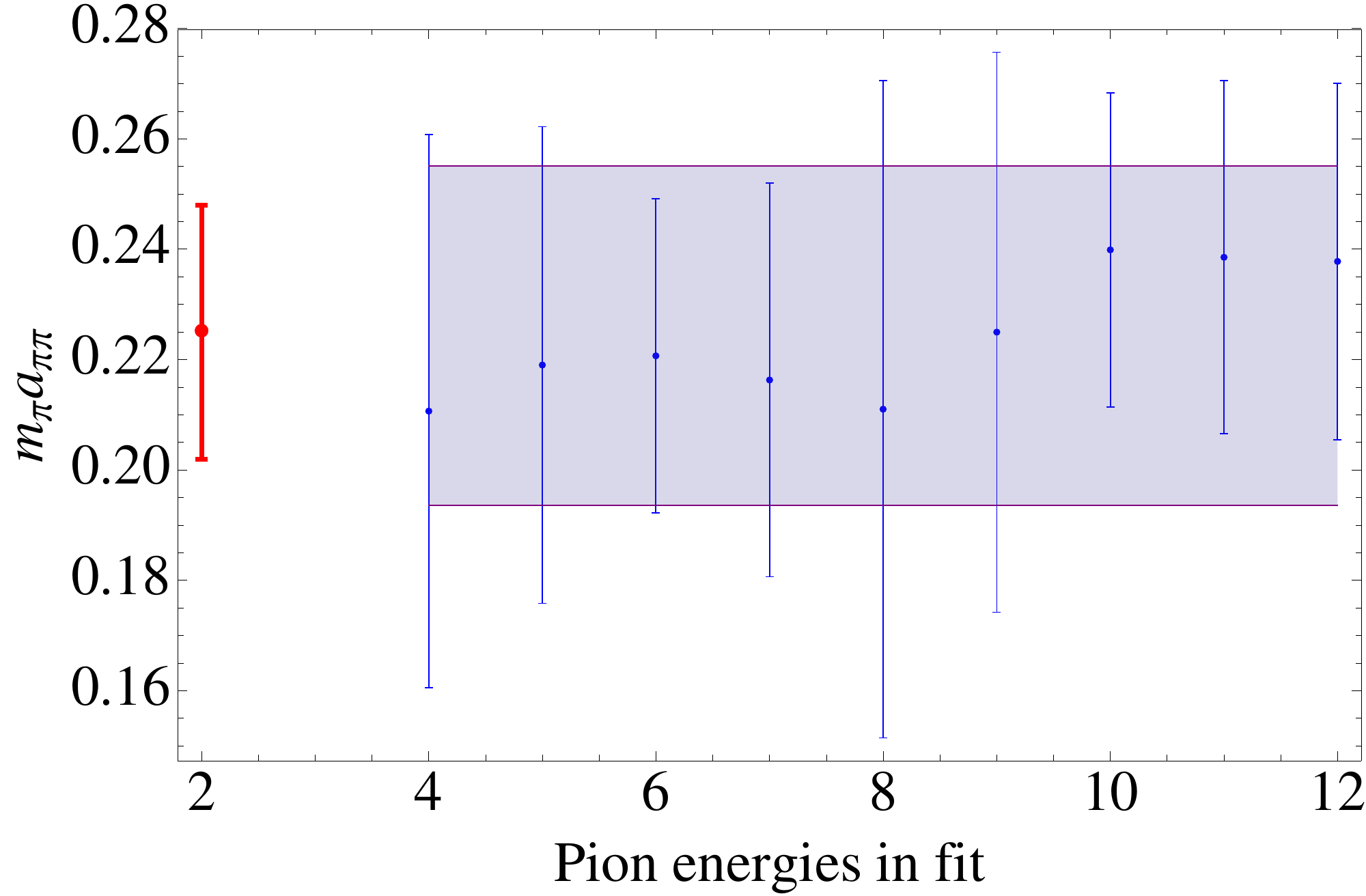} \\
  \includegraphics[scale=0.60,clip=true]{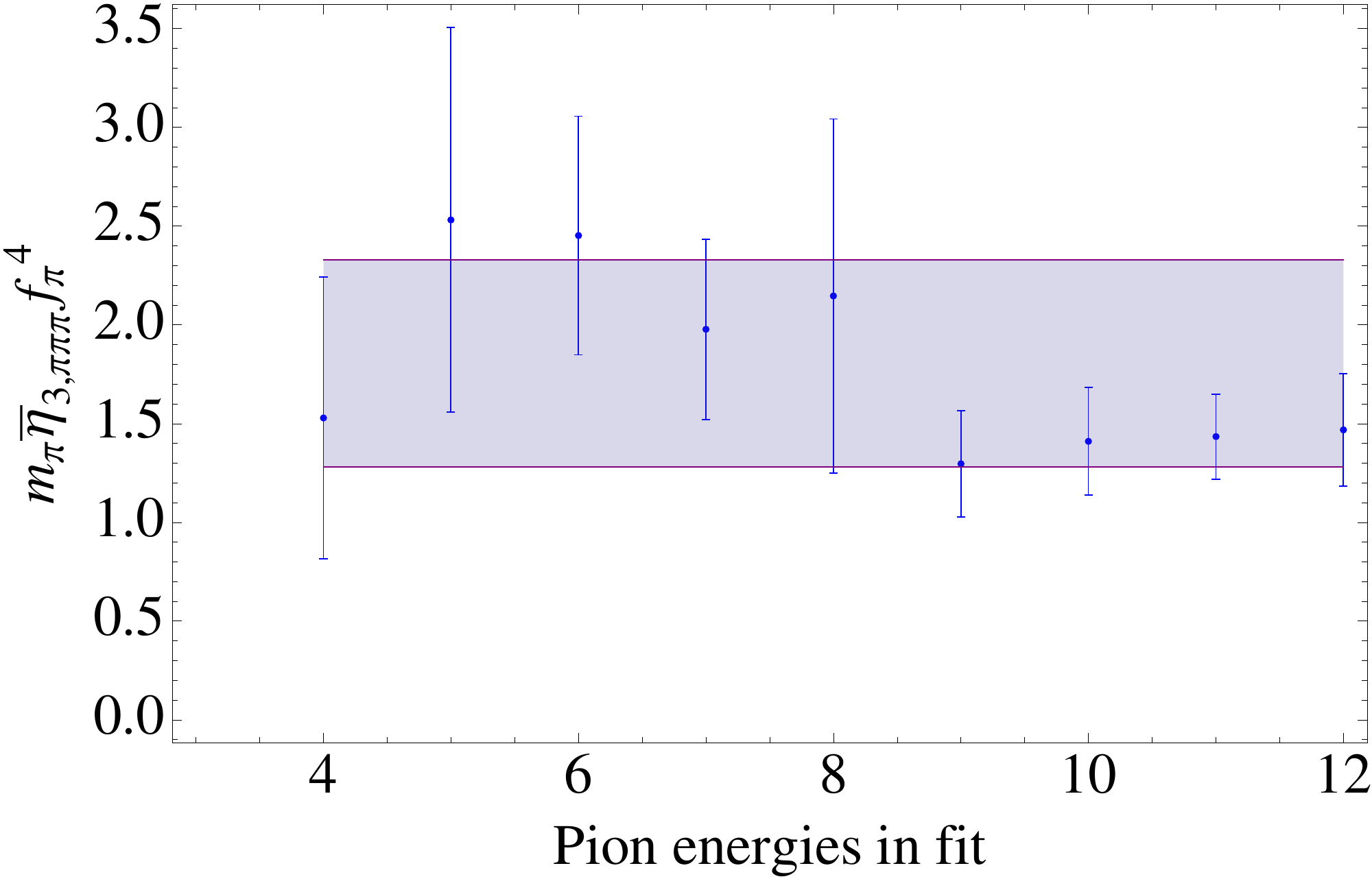}
  \caption{The meaning of the points is identical to that of
   Fig.~\ref{singleMesonScattParams2} only we 
  show pions rather than kaons.}
  \label{singleMesonScattParams3}
\end{figure}

\begin{figure}[htb]
  \centering
  \includegraphics[scale=0.60,clip=true]{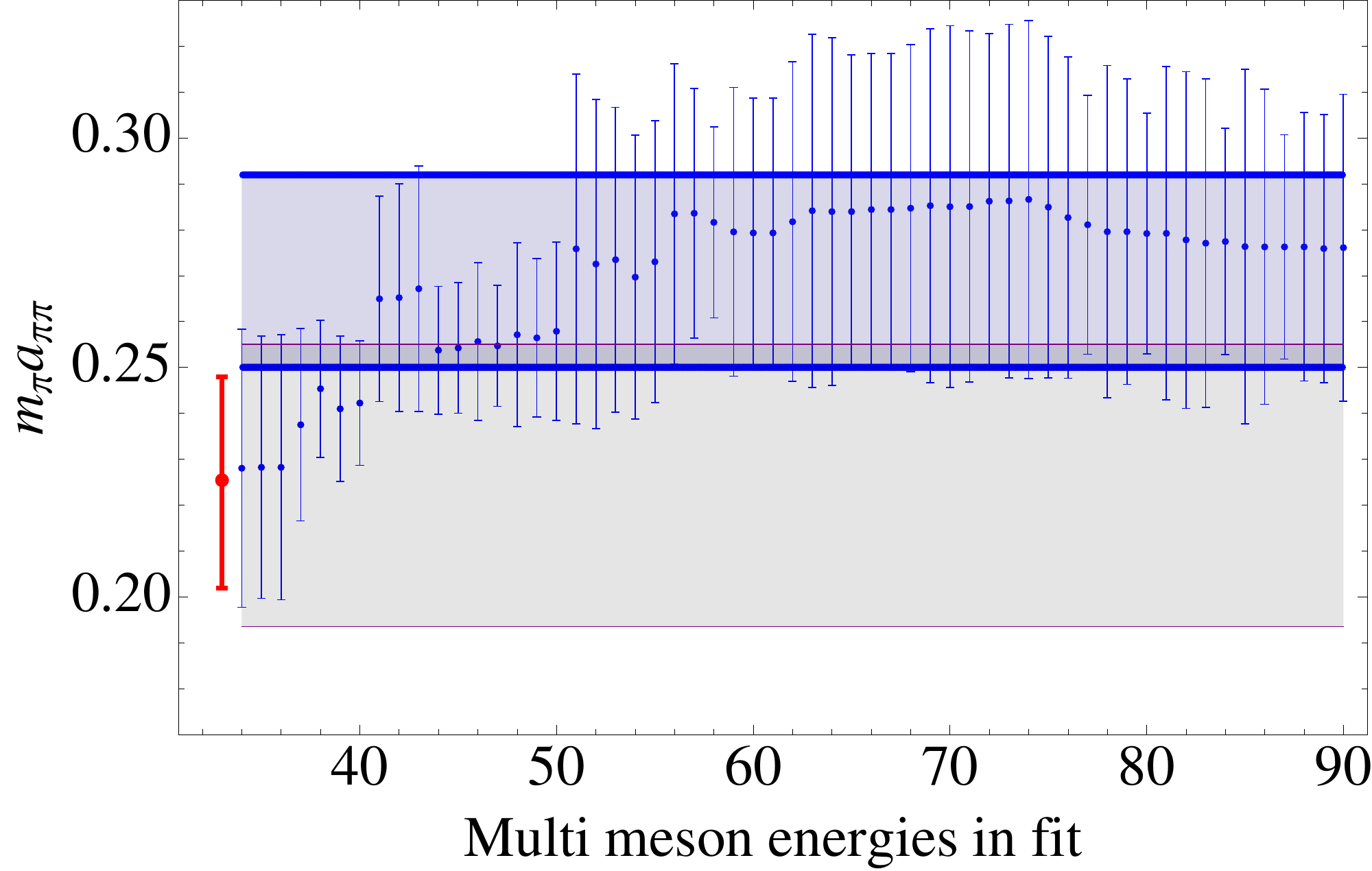} \\
  \includegraphics[scale=0.60,clip=true]{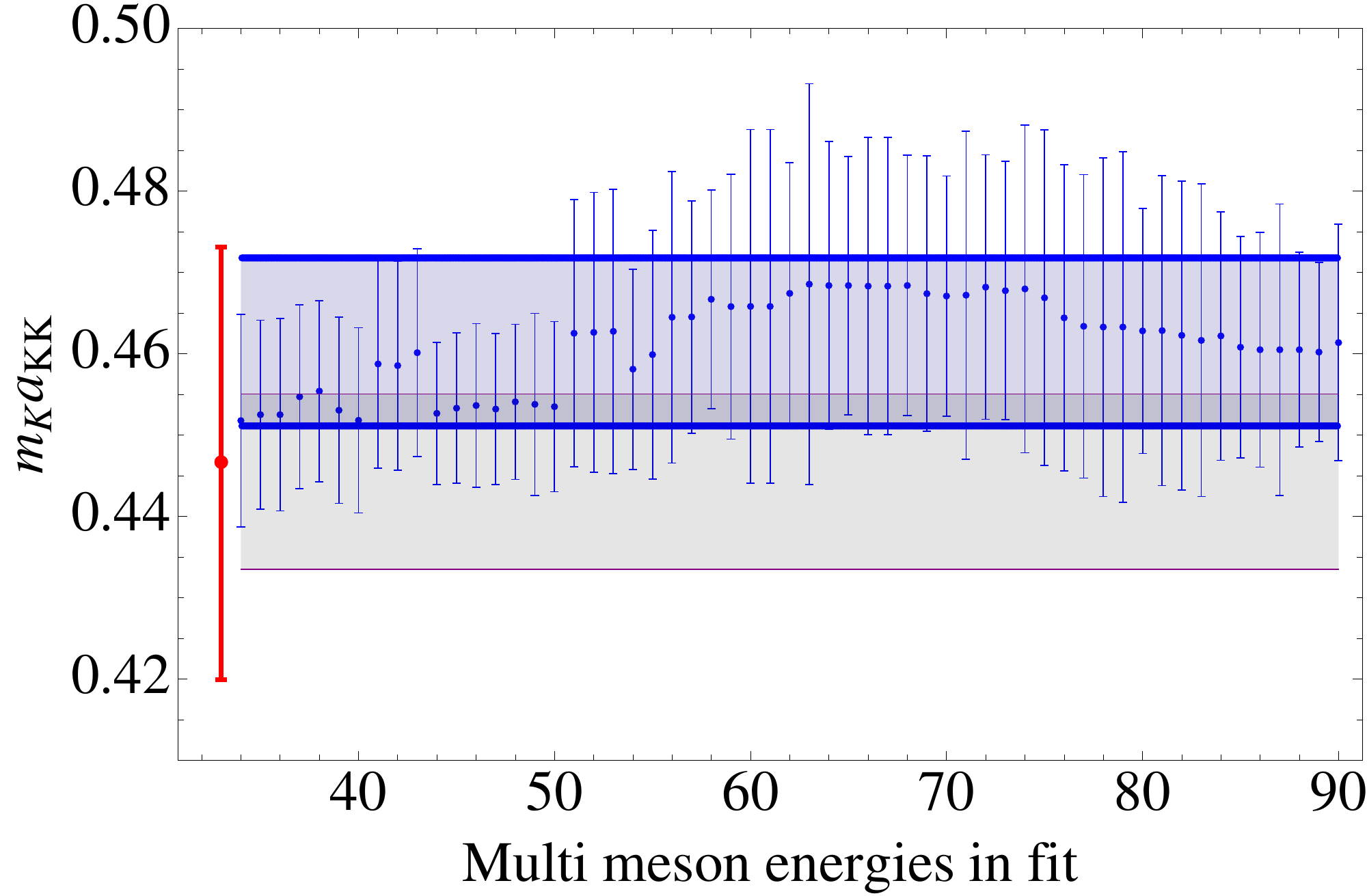} 
  \caption{Calculation of hadronic parameters for mixed-mesons. Extractions are shown
    as a function of the number of energies that were used
    in the extraction. The number of energies used ranged
    from 34 to 90 as discussed in Sec.~\ref{sec:2-3bodyParams}. 
    Uncertainties for a given extraction result from combining statistical 
    and systematic uncertainties in quadrature. The shaded
    band with thick borders denotes the standard deviation of the 
    mean of all extractions in the multi-species case while
    the shaded band with thin borders denotes the standard 
    deviation of the mean of all extractions in the single-species case.
    The red point at the left-most end is the nonperturbative L\"uscher result.}
  \label{multiMesonScattParams1}
\end{figure}

\begin{figure}[htb]
  \centering
  \includegraphics[scale=0.60,clip=true]{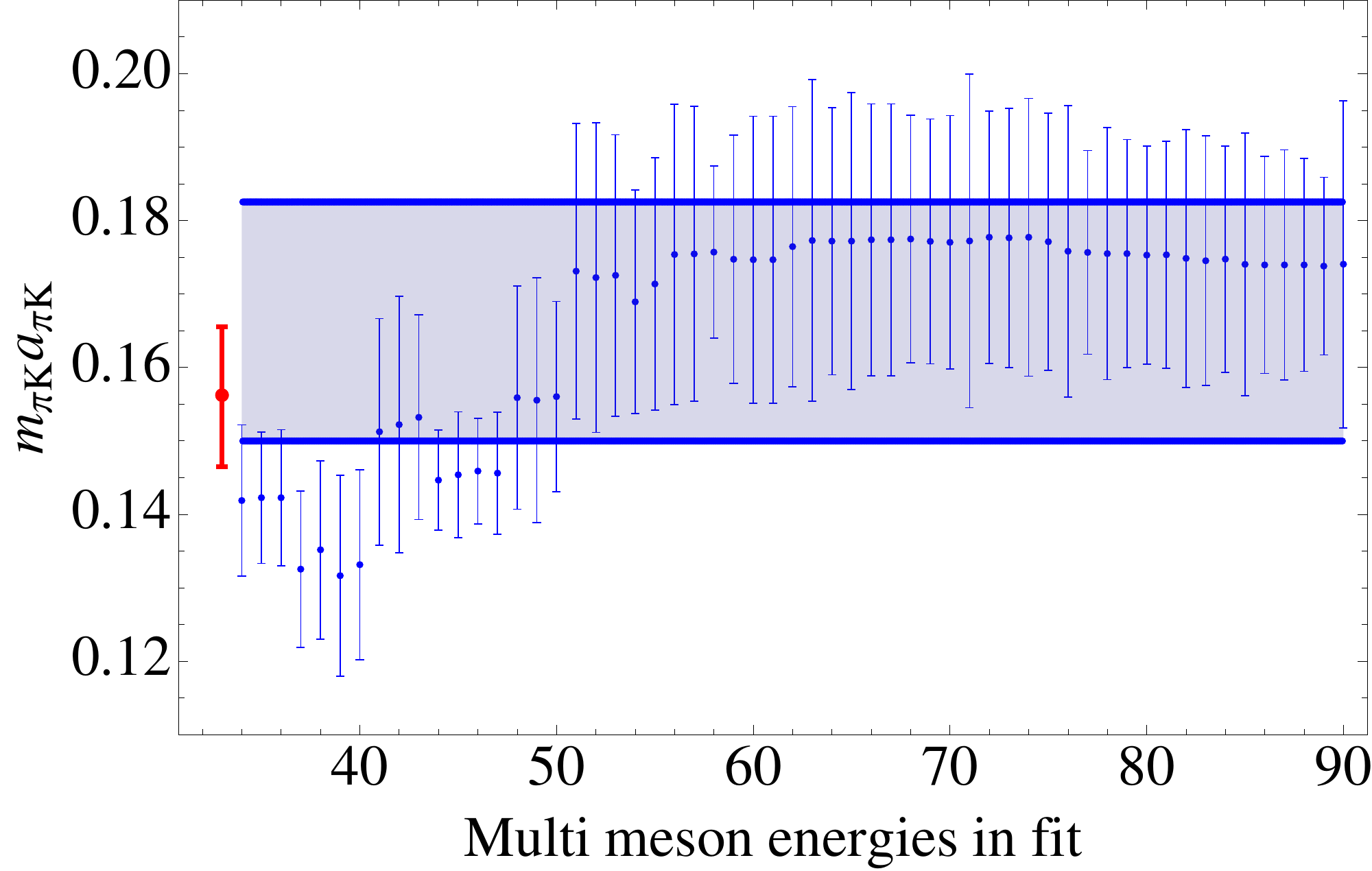} \\
  \includegraphics[scale=0.60,clip=true]{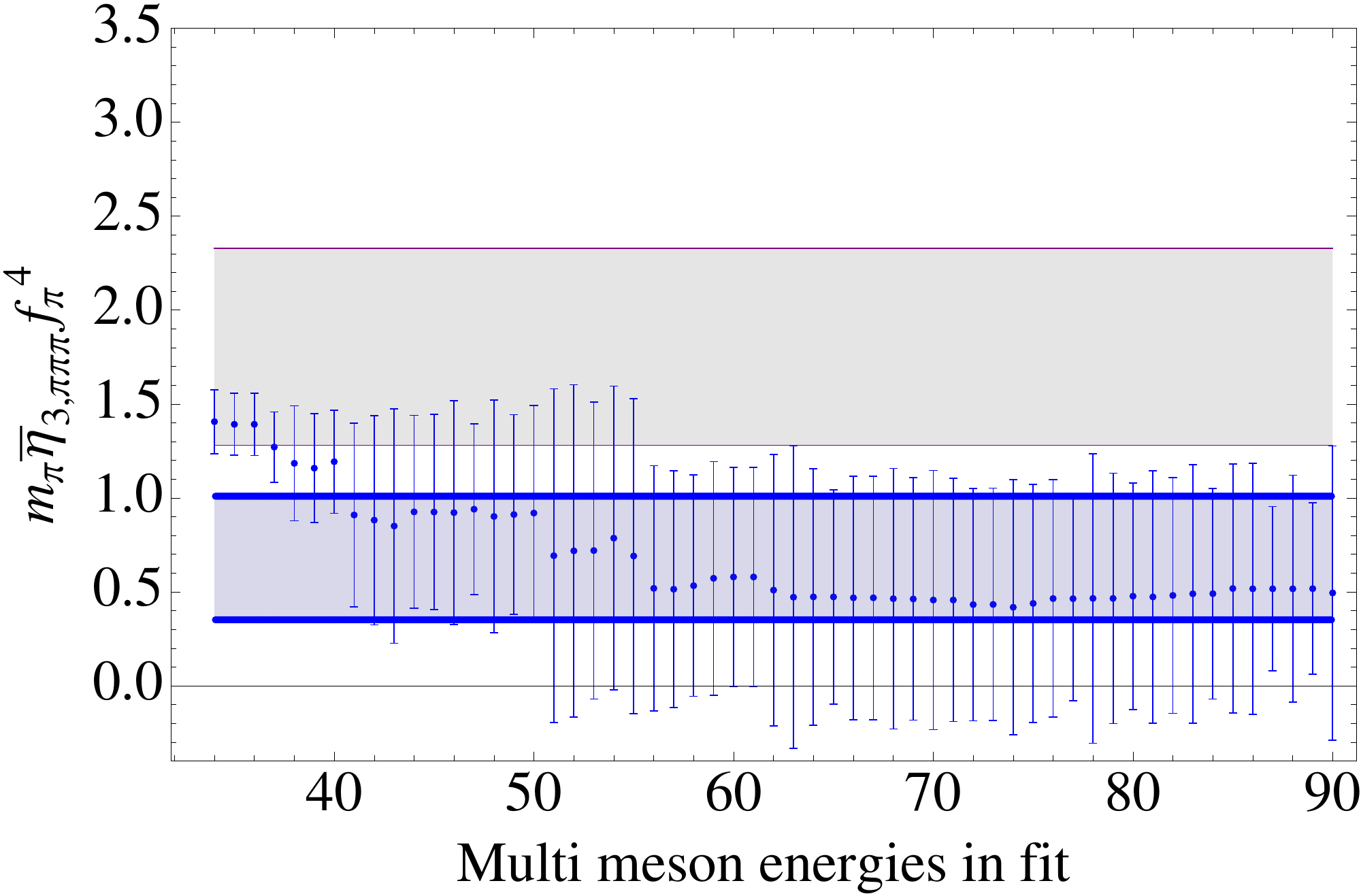}
  \caption{Scattering parameters and three-body interactions
  are shown. The meaning of the points and regions is
  the same as in Fig.~\ref{multiMesonScattParams1}.}
  \label{multiMesonScattParams2}
\end{figure}

\begin{figure}[htb]
  \centering
  \includegraphics[scale=0.60,clip=true]{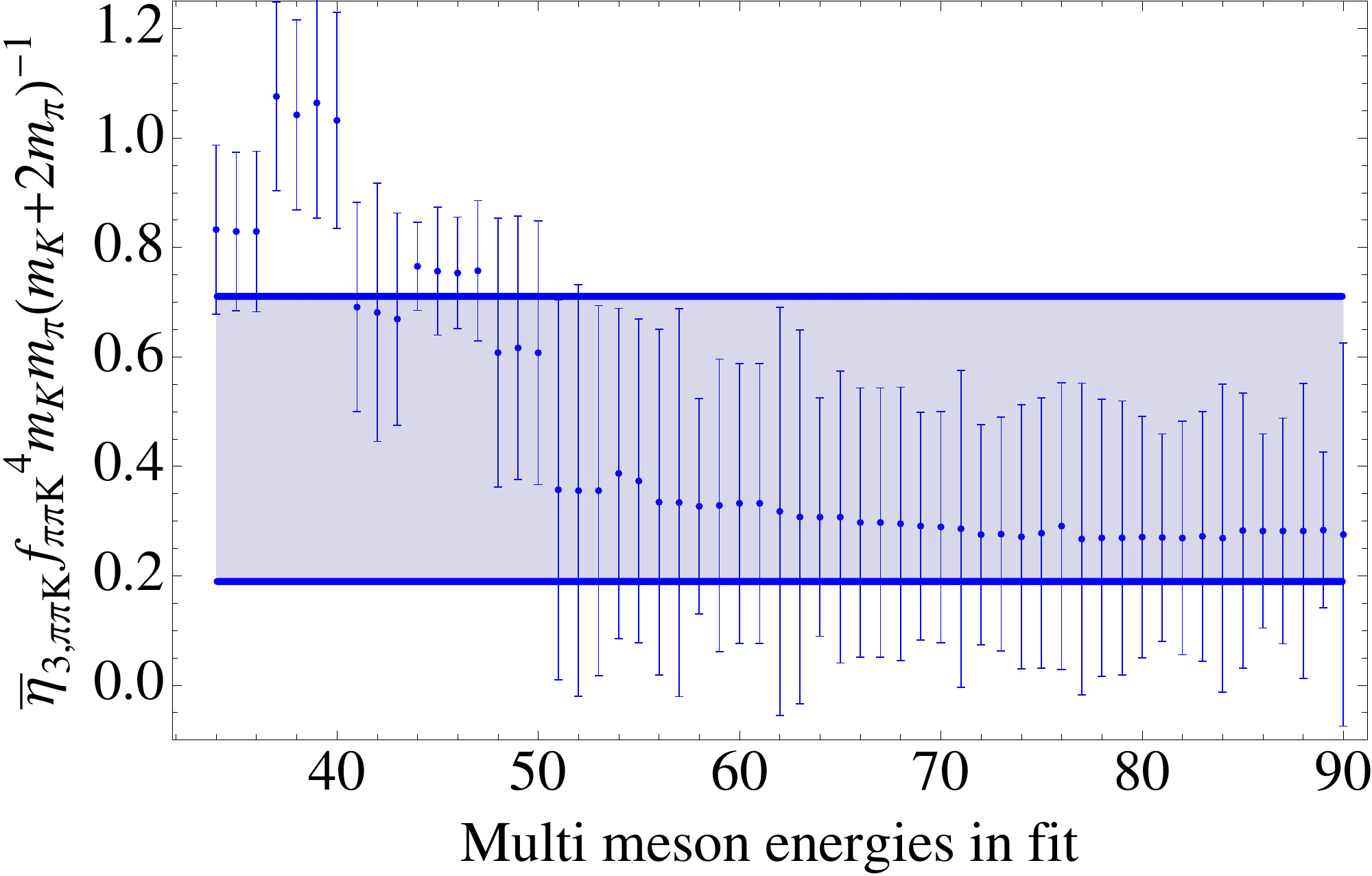} \\
  \includegraphics[scale=0.60,clip=true]{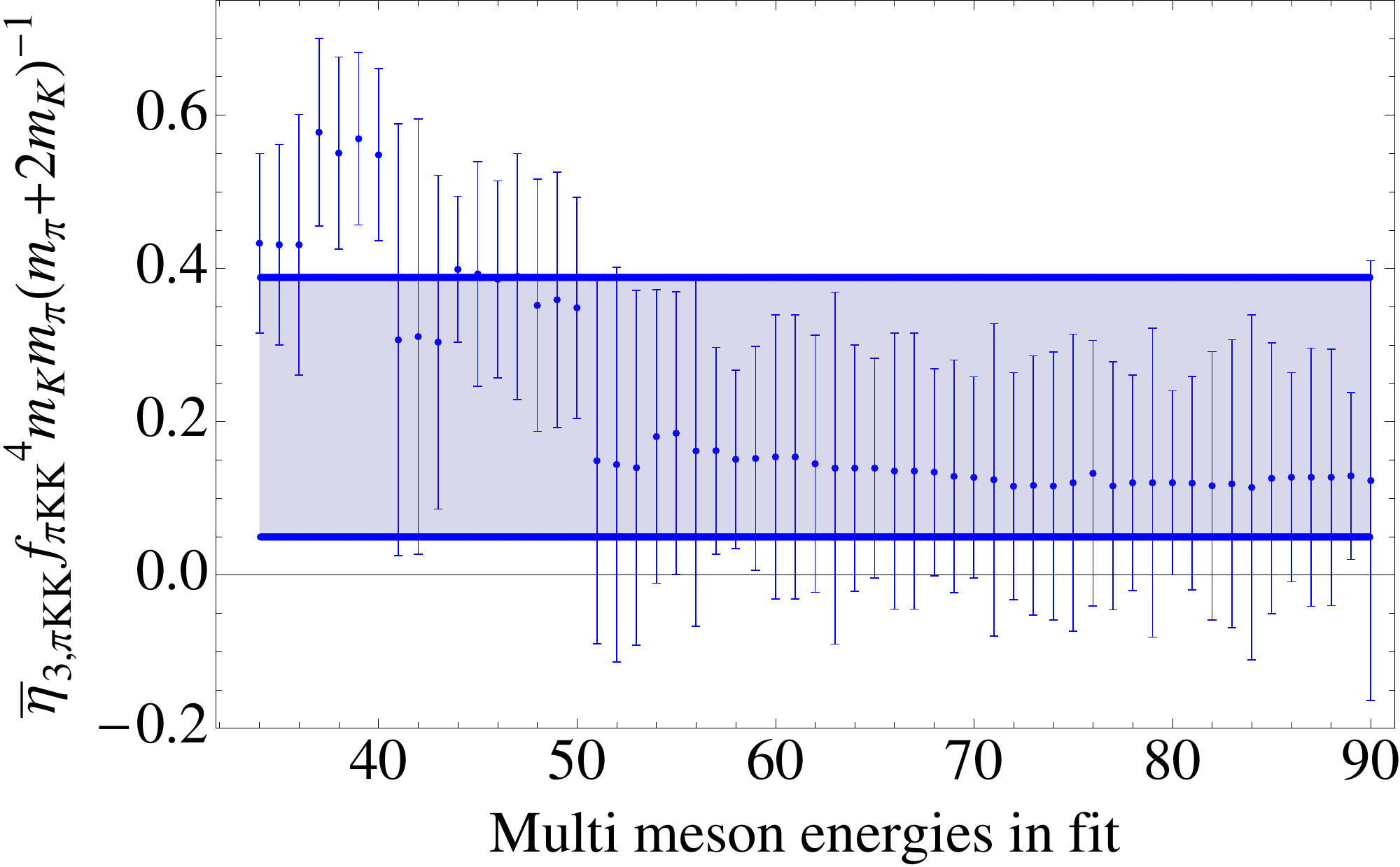}
  \caption{Scattering parameters and three-body interactions
  are shown. The meaning of the points and regions is
  the same as in Fig.~\ref{multiMesonScattParams1}.}
  \label{multiMesonScattParams3}
\end{figure}

The most straightforward determination of the scattering lengths is given by
using the eigenvalue relation from Eq.~(\ref{eq:nonpert-Lusc}).  Using
this, we find,
\begin{align}
  m_{\pi}\bar{a}_{\pi \pi} &= 0.225\pm0.001\pm0.023, \nonumber\\
  m_{K}\bar{a}_{KK} &= 0.4465\pm0.0006\pm0.0266, \nonumber\\
  m_{\pi K}\bar{a}_{\pi K} &= 0.1560\pm0.0004\pm0.0095.
\end{align}
The three-body coefficients can only be determined within the framework of
Eqs.~(\ref{singleSpeciesAnswer}), (\ref{singleSpeciesShift}), and
(\ref{piKanswer}). We also use this same analysis to provide a check
on the above results.  Given that our analysis provides multiple
determinations of the interactions parameters for varying numbers
of combinations of energies used in the fits, these must be combined 
in some way to obtain the final values. Since each separate
extraction can be viewed as a somewhat independent measurement, the
final value given is taken to be the mean from the set of all extractions.  The
final uncertainties on the extractions are combinations of statistical
uncertainties, systematic uncertainties obtained from variation of the
fitting windows as discussed in Sec.~\ref{sec:2-3bodyParams}, and a
second systematic uncertainty determined from the standard deviation
of the full set of extractions, combined in quadrature. The
systematics are the largest source of uncertainties in the
results. The individual extractions of the various parameters and the
final extractions are shown in
Figs.~\ref{singleMesonScattParams2}-\ref{multiMesonScattParams3}.  The
error bars shown combine the statistical, and systematic uncertainties
as discussed in Sec.~\ref{fitStrategies} in quadrature. The shaded
regions with thin borders denote the final results and their
uncertainties. For the mixed species extractions, the second shaded
band with thick borders, denotes the range of uncertainty in the
quoted values from the single species analysis. These are shown
together so the reader can see the overlap region between both sets of
results. The poorest behavior originates from $\bar{\eta}_{3,\pi\pi
  \pi}$ where the mixed species results drift away from
the pure species one.  The final values of the interaction parameters
for the single-species case are:
\begin{align}
  m_{K}\bar{a}_{KK} &= 0.444\pm0.011, \ \ m_{\pi}\bar{a}_{\pi \pi} = 0.224\pm0.031,  \nonumber \\
  m_{K}\bar{\eta}_{3,KKK}f_{K}^4 &= 0.11\pm0.28, \ \
  m_{\pi}\bar{\eta}_{3,\pi\pi\pi}f_{\pi}^4 = 1.81\pm0.52,
\end{align}
whereas for the multi-species case we find:
\begin{align}
  m_{K}\bar{a}_{KK} &= 0.461\pm0.010,  \nonumber\\
  m_{\pi}\bar{a}_{\pi \pi} &= 0.271\pm0.021,  \nonumber\\
  m_{\pi K}\bar{a}_{\pi K} &= 0.166\pm0.016,  \nonumber\\
  m_{K}\bar{\eta}_{3,KKK}f_{K}^4 &=-0.08\pm0.12,  \nonumber\\
  m_{\pi}\bar{\eta}_{3,\pi\pi\pi}f_{\pi}^4 &=0.68\pm0.33, \nonumber \\
  \frac{m_{\pi}m_{K}}{m_{\pi}+2m_K}\bar{\eta}_{3,\pi K K} f_{\pi KK}^4 &= 0.22\pm0.17,  \nonumber\\
  \frac{m_{\pi}m_{K}}{2m_{\pi}+m_K}\bar{\eta}_{3,\pi \pi K} f_{\pi \pi
    K}^4 &= 0.45\pm0.26.
\end{align}
For the two-body parameters, the perturbative analysis (for both pure
species and mixed species) is in agreement with the nonperturbative
values.  In the above results, only the final uncertainty, including
all statistical and systematic contributions, is given as discussed
above. A full list of our extractions can be found in Table~\ref{paramValues}.

\begin{figure}[htb]
  \centering
  \includegraphics[scale=0.60]{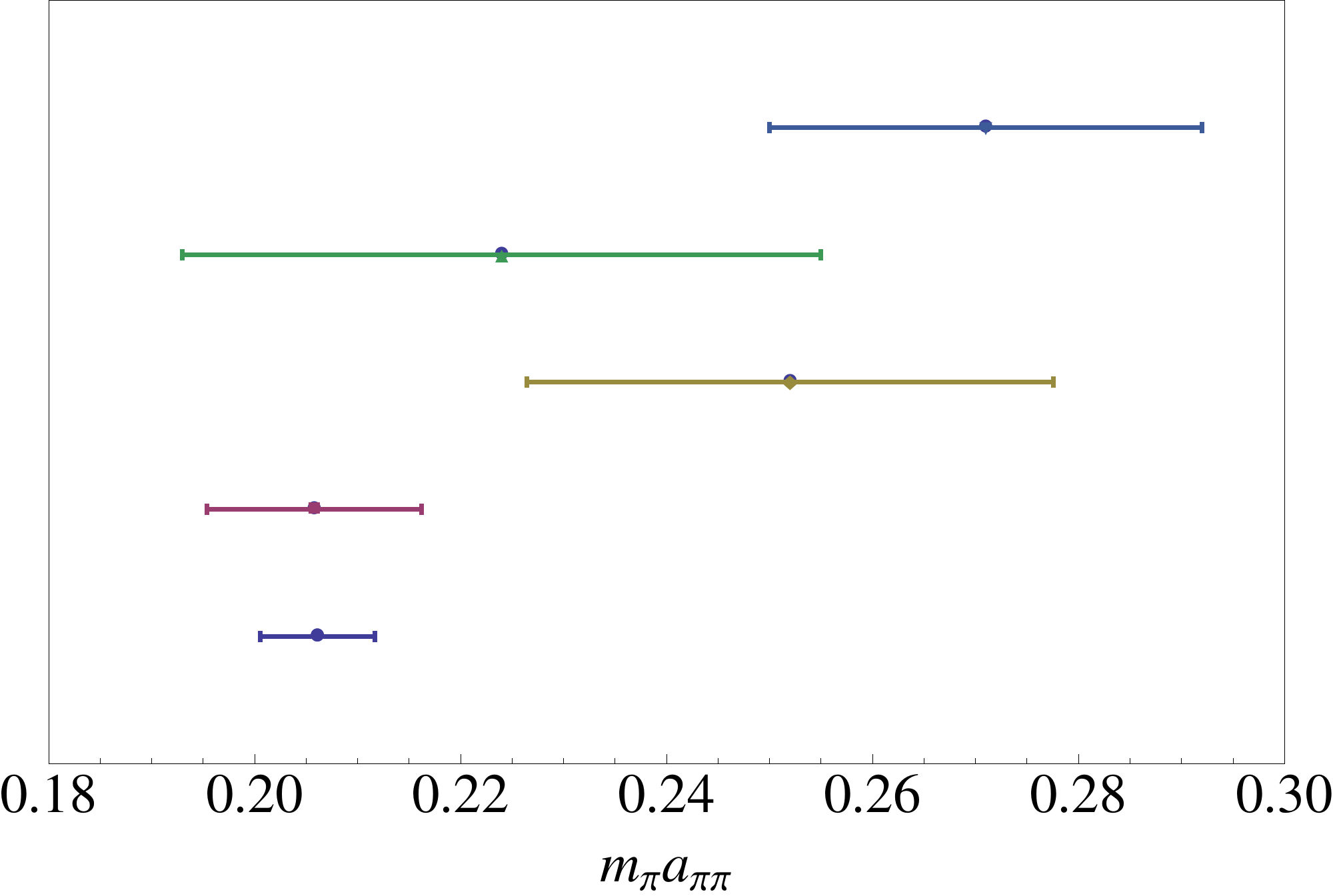}
  \caption{The values of $|m_{\pi} a_{\pi\pi}|$ obtained by different
    groups with pion masses, $m_{\pi} \sim 350 \ \mathrm{MeV}$ are
    shown. From bottom to top, the data are from NPLQCD \cite{Beane:2005rj},
    NPLQCD \cite{Beane:2007es,Detmold:2008fn},
    ETMC \cite{Feng2010268}, the present work's single species value,
    and the present work's multi-species value, respectively. 
    Note that these calculations are at non-zero lattice spacing
    and use different discretizations so complete agreement
    is not expected.}
  \label{collaborationsPionScatLength}
\end{figure}

In previous studies, NPLQCD \cite{Beane:2005rj}, CP-PACS
\cite{Yamazaki:2004qb}, and ETMC \cite{Feng2010268} have measured the
pion scattering lengths in the isospin, $I=2$ channel. The NPLQCD
determination of the $\pi^+ \pi^+$ scattering length at
$m_{\pi}=350 \ \mathrm{MeV}$, yielded $|m_{\pi} a_{\pi\pi}| =
0.2061(49)(17)(20)$ whereas an analysis from multi-pion correlators
\cite{Beane:2007es,Detmold:2008fn} yielded an extraction of $|m_{\pi}
a_{\pi\pi}| = 0.2058(45)^{+46}_{-82}$ and
$m_{\pi}\bar{\eta}_{3,\pi\pi\pi}f_{\pi}^4 = 1.02(08)^{+19}_{-22}$ (in
these results, the first uncertainty is statistical and the other
uncertainties are systematics as discussed in the original
references. In the ETMC analysis
of $\pi^+ \pi^+$ scattering, $|m_{\pi} a_{\pi\pi}| = 0.252(22)(13)$ is
measured at $m_{\pi} \sim 391\ \mathrm{MeV}$. 
Fig.~\ref{collaborationsPionScatLength} shows the dimensionless
combination $|m_{\pi}a_{\pi \pi}|$ from the current work in comparison
to the determinations by other groups at a similar pion mass. However,
these results are at non-zero lattice spacing and correspond to
different discretizations, so agreement is not necessary.  The $I=1$,
$K^+ K^+$ scattering length was also determined by the NPLQCD
collaboration in Ref.~\cite{Beane:2007uh}. An analysis of the
two-point kaon correlator yielded a value of $|m_{K} a_{KK}| =
0.497(10)(22)$, again at $m_\pi\sim 350$ MeV. Analysis of multi-kaon
correlators \cite{Detmold:2008yn}, led to $|m_{K} a_{KK}| =
0.503(11)(19)$ and $m_{K}\bar{\eta}_{3,KKK}f_{K}^4 =-0.1(2)(5)$ where
the uncertainties are statistical and systematic respectively.  The
$\pi K$ scattering length has been investigated in quenched
\cite{Miao:2004gy} and full QCD \cite{Beane:2006gj} and the unquenched
determination at $m_\pi\sim 350$ MeV is $m_{\pi K}\bar{a}_{\pi
  K}=0.155(40)$.  Hence, it is clear the current results are generally
consistent with other groups' extractions. The mixed species
three-body parameters are novel results and are found to be of
natural size and positive.

\subsection{Isospin and Hypercharge Chemical Potentials}
\label{sec:isosp-hyperch-chem}

As we have determined the dependence of the energy of the mixed meson
systems on the number of pions and kaons, we can construct the isospin
and hypercharge (or strangeness) chemical potentials using finite
differences following Refs.~\cite{Detmold:2008yn, Detmold:2008fn}
where systems of pions and kaons were investigated separately. In
Refs.~\cite{Detmold:2008yn, Detmold:2008fn}, remarkable
agreement was found between the numerical results
and the leading order (LO) $\chi$PT
prediction \cite{Son:2000by} of the relation between the
isospin(hypercharge) density and chemical potential
\begin{equation}
  \rho_j = \frac{f_{j}^2 \mu_{j}}{2} \Big(1-\frac{m_{j}^4}{\mu_{j}^4}  \Big),
\end{equation}
with $j \in (\pi,K)$.  The situation here is more
complicated since there are finite differences acting in various
non-orthogonal directions; the differences between $E_{N_\pi, N_K}$ and
$E_{N_\pi-1, N_K}$ determine $\mu_I$ while  linear combinations of
$E_{N_\pi, N_K}, E_{N_\pi-1, N_K}$, and $E_{N_\pi, N_K-1}$ determine
$\mu_S$\footnote{$\mu_{K^+} = \mu_S + \mu_I /2$ with
  $\mu_{K^+}=E_{N_\pi, N_K}-E_{N_\pi, N_K-1}$ and $\mu_{I}=E_{N_\pi,
    N_K}-E_{N_\pi-1, N_K}$}.  One goal of this analysis is to see
where on the $\mu_S$ vs. $\mu_I$ phase diagram \cite{Kogut:2001id},
the states created in the lattice calculation lie.

In Ref.~\cite{Kogut:2001id}, leading order SU(3)
$\chi$PT is used to predict three distinct phases for nonzero isospin
 and hypercharge chemical potential. The first is the normal
phase where the ground state has a net particle number of
zero. The other two phases are the pion-condensed and kaon-condensed
phases. The transition between the kaon-condensed phase and the 
pion-condensed phase is predicted to be a
first order phase transition, separated by the line $\mu_S =
\left(-m_{\pi}^2+\sqrt{(m_{\pi}^2-\mu_I^2)^2 +4m_K^2\mu_I^2}\right)/2\mu_I$,
while the transition from the normal phase to either 
condensed phase is expected to be of second order \footnote{An
AdS/QCD based model \cite{Albrecht:2010eg} finds these transitions to be of
first order.} and are defined by the lines $\mu_S=m_K-\mu_I/2$ and $\mu_I =
m_{\pi}$. These predictions assume zero temperature and are likely softened 
by the non-zero temperature at which the lattice calculation is performed \cite{Loewe:2002tw}.

\begin{figure}[htb]
  \centering
  \includegraphics[scale=0.9,clip=true]{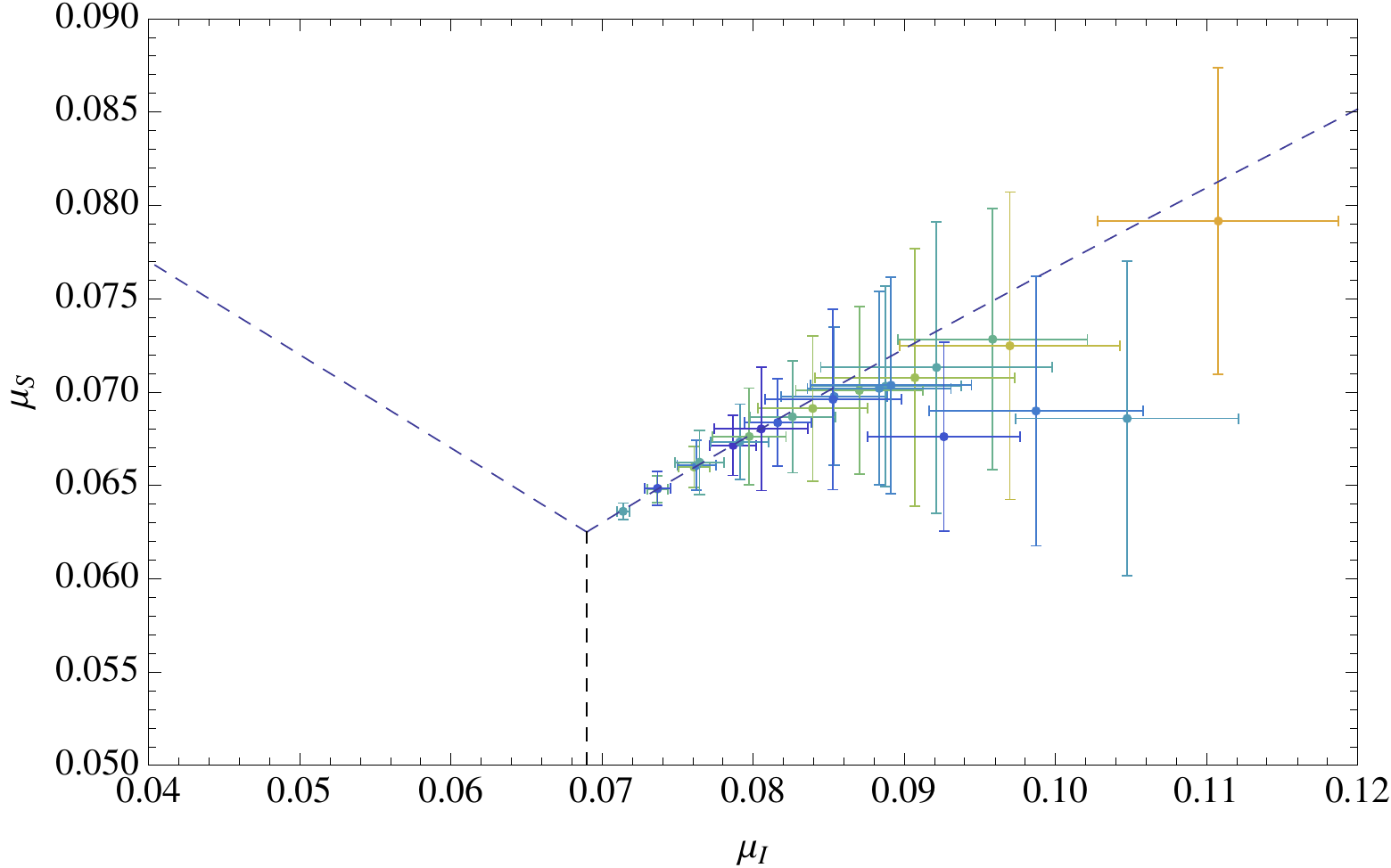}
  \caption{  \label{phasePlot}
$\mu_S$ vs. $\mu_I$. The data points corresponds to the lattice
    calculations, colored by the energy of the system from low (blue) to
    high (orange). The dashed lines are predictions of $\chi$PT. The
    lower-left region is the normal phase, the right-hand region is
    the pion- condensed phase, will the upper portion is the kaon-
    condensed phase.}
\end{figure}
In Fig.~\ref{phasePlot}, both the lattice calculations of $(\mu_I,\mu_S)$ and 
the $\chi$PT phase boundaries are shown (dashed lines). 
Data points corresponding to higher numbers of
particle states are shown in a orange/reddish color, while lower
numbers are given in a blueish/greenish color. Points with large uncertainties 
are excluded from this figure for clarity (the omitted data 
correspond to the highest particle numbers).
It is striking that the calculated chemical potentials mostly lie near the
first-order phase transition line predicted by $\chi$PT. Further calculations with larger
numbers of pions and kaons  will be enlightening, but more complex probes of these 
systems may be needed to fully understand the states that have been produced.

\section{Conclusions}
\label{sec:disc-and-conc}

In this work, we have numerically studied complex systems 
of mesons of two distinct flavors, like-charged pions and  
kaons, and used them to extract information about the two- 
and three- body interactions amongst pions and kaons. 
Where known, the interactions were found to be consistent 
with previous calculations, however, two mixed-species 
three-body interactions were determined for the first time. 
Additionally, the isospin and strangeness chemical potentials 
and phase structure of the system have been investigated, 
with the systems preferring to probe a region in the 
$(\mu_I,\mu_s)$ plane where $\chi$PT predicts a first 
order phase transition.

A major aim of this work was to investigate technical 
issues that arise in the analysis of complex multi-hadron 
systems. Accounting for the thermal states that proliferate 
in such systems, which easily factorize into distinct color 
singlet states, proved challenging and future calculations 
should avoid this by using larger temporal extents. 
Additionally, a  number of techniques to perform coupled 
fits to the ${\cal O}(100)$ correlators studied were investigated 
and found to be beneficial in the analysis.

In the future, calculations probing larger meson numbers 
will allow further investigations of the phase structure of 
these interesting QCD systems. To understand the 
structure of the condensed systems created in the current 
and future calculations, more complicated observables that
access transport properties may be needed; investigations 
in this direction are under consideration.

\acknowledgments We would like to thank 
C. Aubin, J. Erlich, K. Orginos, A. Nicholson, M. Savage, 
Z. Shi, A. Walker-Loud, and J. Wasem for comments and 
many useful conversations, and the NPLQCD and Hadron 
Spectrum collaborations for the use of their propagators 
and gauge configurations, respectively.  
BS would like to acknowledge the NSC of R.O.C.
WD is supported in part by US Department of Energy grants 
DE-AC05-06OR23177 (JSA) and DE-FG02-04ER41302
and by OJI grant DE-SC0001784 and Jeffress Memorial Trust grant J-968. 
Calculations made use of computational resources provided by the 
NSF Teragrid and DOE NERSC facility and local resources at the 
University of Washington and the College of William and Mary.

\newpage

\appendix
\section{Tables}
\label{tables}

\begin{table}[htb]\label{energyTable}
  \footnotesize
  \caption{Extraction of single-species meson energies (in lattice units). The
    first set of uncertainties are statistical and the second
    set are systematic uncertainties associated with shifting the correlator
    fitting window by $\Delta t=\pm 1$. The remaining column shows
    the fit range chosen.} 
  \label{singleSpeciesEnergyTable}
  
  \centering      
  \begin{tabular}{|  c  ||  l    c   |}  
    \hline
    $N,M$ & $\langle{E}\rangle \pm \sigma_{\mathrm{stat}} \pm \sigma_{\mathrm{sys}}$ & 
    $(t_{\mathrm{min}},t_{\mathrm{max}})$ \\ [2.0ex] 
    \hline\hline
    $1,0$ & $0.06936\pm0.00021\pm0.00002$ & (27,37)\\

    $2,0$ & $0.14067\pm0.00049\pm0.00007$ & (27,37)\\

    $3,0$ & $0.2143\pm0.0010\pm0.0004$ & (27,37)\\

    $4,0$ & $0.291\pm0.002\pm0.001$ & (27,37) \\

    $5,0$ & $0.378\pm0.003\pm0.003$ & (31,41)\\

    $6,0$ & $0.470\pm0.005\pm0.004$ & (31,41)\\

    $7,0$ & $0.547\pm0.005\pm0.003$ & (26,36)\\

    $8,0$ & $0.654\pm0.007\pm0.003$ & (31,41)\\

    $9,0$ & $0.7560\pm0.0090\pm0.0007$ & (31,41) \\

    $10,0$ & $0.868\pm0.014\pm0.002$ & (31,41)\\

    $11,0$ & $0.983\pm0.015\pm0.003$ & (32,42)\\

    $12,0$ & $1.101\pm0.046\pm0.009$ & (25,35)\\

    $0,1$ & $0.09727\pm0.00015\pm0.00001$ & (27,37)\\

    $0,2$ & $0.19663\pm0.00031\pm0.00007$ & (27,37) \\

    $0,3$ & $0.2982\pm0.0005\pm0.0004$ & (27,37) \\

    $0,4$ & $0.4021\pm0.0008\pm0.0011$ & (27,37) \\

    $0,5$ & $0.5083\pm0.0011\pm0.0008$ & (27,37) \\

    $0,6$ & $0.617\pm0.002\pm0.003$ & (27,37)\\

    $0,7$ & $0.728\pm0.002\pm0.007$ & (28,39) \\

    $0,8$ & $0.842\pm0.003\pm0.004$ & (28,39) \\

    $0,9$ & $0.959\pm0.005\pm0.011$ & (28,39) \\

    $0,10$ & $1.101\pm0.036\pm0.009$ & (32,49) \\

    $0,11$ & $1.23\pm0.03\pm0.01$ & (27,41)\\

    $0,12$ & $1.384\pm0.076\pm0.001$ & (33,52) \\[2ex] 
    \hline\hline
  \end{tabular}
  \label{table:q}  
\end{table}

\begin{center}
  \begin{longtable}{| l || l c |}
    \caption{Extraction of multi-species meson energies (in lattice
      units). The columns are the same as in Table
      \ref{singleSpeciesEnergyTable}.} 
    \label{multiSpeciesEnergyTable} \\

    \hline \multicolumn{1}{|c|}{$N,M$} & \multicolumn{1}{c|}{$\langle{E}\rangle \pm \sigma_{\mathrm{stat}} \pm \sigma_{\mathrm{sys}}$} & \multicolumn{1}{c|}{$(t_{\mathrm{min}},t_{\mathrm{max}})$} \\
    \hline\hline
    \endfirsthead

    \multicolumn{3}{c}%
    {{\bfseries \tablename\ \thetable{} -- continued from previous page}} \\
    \hline \multicolumn{1}{c|}{$N,M$} &
    \multicolumn{1}{c|}{$\langle{E}\rangle \pm \sigma_{\mathrm{stat}}
      \pm \sigma_{\mathrm{sys}}$} &
    \multicolumn{1}{c|}{$(t_{\mathrm{min}},t_{\mathrm{max}})$} \\
    \hline\hline
    \endhead

    \hline \multicolumn{3}{|r|}{{Continued on next page}} \\ \hline
    \endfoot

    \hline \hline
    \endlastfoot

    $1,1$ & $0.1687\pm0.0004\pm0.0002$ & (27,37)\\

    $1,2$ & $0.2703\pm0.0006\pm0.0006$ & (27,37)\\

    $1,3$ & $0.3743\pm0.0009\pm0.0013$ & (27,37) \\

    $1,4$ & $0.481\pm0.001\pm0.002$ & (27,37) \\

    $1,5$ & $0.590\pm0.002\pm0.004$ & (27,37)\\

    $1,6$ & $0.7024\pm0.0032\pm0.0005$ & (28,39) \\

    $1,7$ & $0.8171\pm0.0043\pm0.0003$ & (28,39) \\

    $1,8$ & $0.938\pm0.005\pm0.005$ & (27,37) \\

    $1,9$ & $1.07\pm0.01\pm0.12$ & (31,49) \\

    $1,10$ & $1.21\pm0.02\pm0.19$ & (27,41) \\

    $1,11$ & $1.36\pm0.08\pm0.05$ & (33,49) \\

    $2,1$ & $0.2423\pm0.0008\pm0.0005$ & (27,37) \\

    $2,2$ & $0.347\pm0.001\pm0.001$ & (27,37) \\

    $2,3$ & $0.453\pm0.002\pm0.002$ & (27,37) \\

    $2,4$ & $0.563\pm0.002\pm0.003$ & (27,37) \\

    $2,5$ & $0.6770\pm0.0037\pm0.0007$ & (28,39) \\

    $2,6$ & $0.793\pm0.006\pm0.003$ & (28,39) \\

    $2,7$ & $0.914\pm0.006\pm0.007$ & (27,37)\\

    $2,8$ & $1.049\pm0.006\pm0.057$ & (31,49)\\

    $2,9$ & $1.19\pm0.01\pm0.04$ & (33,50) \\

    $2,10$ & $1.33\pm0.04\pm0.01$ & (33,49) \\

    $3,1$ & $0.319\pm0.001\pm0.001$ & (27,37) \\

    $3,2$ & $0.426\pm0.002\pm0.002$ & (27,37) \\

    $3,3$ & $0.537\pm0.003\pm0.002$ & (27,37) \\

    $3,4$ & $0.652\pm0.005\pm0.002$ & (28,39) \\

    $3,5$ & $0.769\pm0.007\pm0.004$ & (28,39) \\

    $3,6$ & $0.889\pm0.008\pm0.003$ & (28,38) \\

    $3,7$ & $1.02\pm0.01\pm0.10$ & (27,38) \\

    $3,8$ & $1.16\pm0.03\pm0.01$ & (33,48) \\

    $3,9$ & $1.305\pm0.040\pm0.006$ & (33,49) \\

    $4,1$ & $0.400\pm0.003\pm0.001$ & (27,37) \\

    $4,2$ & $0.512\pm0.004\pm0.001$ & (27,37) \\

    $4,3$ & $0.626\pm0.004\pm0.002$ & (28,39) \\

    $4,4$ & $0.745\pm0.007\pm0.005$ & (28,39) \\

    $4,5$ & $0.868\pm0.011\pm0.008$ & (28,39) \\

    $4,6$ & $0.99\pm0.01\pm0.05$ & (27,38) \\

    $4,7$ & $1.132\pm0.017\pm0.009$ & (33,47) \\

    $4,8$ & $1.280\pm0.03\pm0.001$ & (33,49) \\

    $5,1$ & $0.492\pm0.004\pm0.004$ & (31,41) \\

    $5,2$ & $0.610\pm0.006\pm0.004$ & (31,41) \\

    $5,3$ & $0.7313\pm0.0061\pm0.0007$ & (31,42) \\

    $5,4$ & $0.856\pm0.010\pm0.001$ & (31,42) \\

    $5,5$ & $0.982\pm0.01\pm0.007$ & (31,43) \\

    $5,6$ & $1.113\pm0.023\pm0.002$ & (33,46) \\

    $5,7$ & $1.250\pm0.028\pm0.003$ & (33,47) \\

    $6,1$ & $0.590\pm0.007\pm0.003$ & (31,41) \\

    $6,2$ & $0.697\pm0.006\pm0.007$ & (27,37) \\

    $6,3$ & $0.837\pm0.009\pm0.002$ & (31,42) \\

    $6,4$ & $0.968\pm0.012\pm0.002$ & (31,41) \\

    $6,5$ & $1.103\pm0.025\pm0.006$ & (33,43) \\

    $6,6$ & $1.2349\pm0.0244\pm0.0005$ & (33,47) \\

    $7,1$ & $0.664\pm0.006\pm0.007$ & (26,36) \\

    $7,2$ & $0.8098\pm0.009\pm0.0003$ & (33,43) \\

    $7,3$ & $0.949\pm0.011\pm0.003$ & (31,41) \\

    $7,4$ & $1.091\pm0.016\pm0.002$ & (32,42) \\

    $7,5$ & $1.19\pm0.05\pm0.02$ & (28,38) \\

    $8,1$ & $0.770\pm0.006\pm0.009$ & (27,37) \\

    $8,2$ & $0.914\pm0.010\pm0.003$ & (31,41) \\

    $8,3$ & $1.071\pm0.015\pm0.002$ & (32,42) \\

    $8,4$ & $1.160\pm0.051\pm0.027$ & (28,38) \\

    $9,1$ & $0.8757\pm0.0084\pm0.0002$ & (31,42) \\

    $9,2$ & $1.034\pm0.014\pm0.002$ & (32,42) \\

    $9,3$ & $1.13\pm0.05\pm0.03$ & (28,38) \\

    $10,1$ & $0.988\pm0.011\pm0.002$ & (32,42) \\

    $10,2$ & $1.11\pm0.04\pm0.03$ & (28,38) \\

    $11,1$ & $1.105\pm0.014\pm0.005$ & (32,42) \\

  \end{longtable}
\end{center}

Letting:
\begin{align}
  F_1 &= m_{K} \bar{a}_{KK}, \ \ F_2 = m_{\pi} \bar{a}_{\pi\pi}, \ \ F_3 = m_{\pi K} \bar{a}_{\pi K}, \nonumber\\
  F_4 &= m_{K} \bar{\eta}_{3,K} f_K^4, \ \ F_5 = m_{\pi} \bar{\eta}_{3,\pi} f_{\pi}^4, \nonumber\\
  F_6 &= \frac{m_{\pi}m_{K}\bar{\eta}_{3,\pi K K} f_{\pi KK}^4}{m_{\pi}+2m_K}, \ \ F_7 =
  \frac{m_{\pi}m_{K}\bar{\eta}_{3,\pi \pi K} f_{\pi \pi K}^4}{2m_{\pi}+m_K}. \nonumber
\end{align}

\begin{center}
  \begin{longtable}{| c | l | l | l | l | l | l | l |}
    \caption{Results for scattering and three-body parameters in
      lattice units. The uncertainties shown are statistical and systematic,
      respectively.}
    \label{paramValues} \\

    \hline \multicolumn{1}{|c|}{Num} & \multicolumn{1}{c|}{$F_1$} &
    \multicolumn{1}{c|}{$F_2$} & \multicolumn{1}{c|}{$F_3$} &
    \multicolumn{1}{c|}{$F_4$} & \multicolumn{1}{c|}{$F_5$} &
    \multicolumn{1}{c|}{$F_6$} & \multicolumn{1}{c|}{$F_7$} \\ \hline
    \endfirsthead

    \multicolumn{8}{c}%
    {{\bfseries \tablename\ \thetable{} -- continued from previous page}} \\
    \hline \multicolumn{1}{c|}{Num} & \multicolumn{1}{c|}{$F_1$} &
    \multicolumn{1}{c|}{$F_2$} & \multicolumn{1}{c|}{$F_3$} &
    \multicolumn{1}{c|}{$F_4$} & \multicolumn{1}{c|}{$F_5$} &
    \multicolumn{1}{c|}{$F_6$} & \multicolumn{1}{c|}{$F_7$} \\ \hline
    \endhead

    \hline \multicolumn{8}{|r|}{{Continued on next page}} \\ \hline
    \endfoot

    \hline \hline
    \endlastfoot

    34 & 0.451(10)(7) & 0.22(1)(2) & 0.141(9)(4) & 0.046(21)(3) & \
    1.4(16)(3) & 0.43(10)(5) & 0.83(14)(5)\\

    35 & 0.452(10)(4) & 0.22(1)(2) & 0.1422(89)(7) & 0.056(21)(1) & \
    1.39(165)(8) & 0.43(10)(8) & 0.83(14)(0)\\

    36 & 0.452(10)(4) & 0.22(1)(2) & 0.142(8)(2) & 0.056(22)(0) & \
    1.3(16)(1) & 0.4(1)(1) & 0.82(14)(2)\\

    37 & 0.455(11)(0) & 0.23(1)(1) & 0.133(11)(0) & 0.1(2)(1) & \
    1.2(16)(9) & 0.57(11)(2) & 1.(15)(6)\\

    38 & 0.455(11)(0) & 0.245(14)(2) & 0.135(10)(5) & 0.1(2)(2) & \
    1.(1)(2) & 0.55(12)(1) & 1.(15)(6)\\

    39 & 0.453(11)(2) & 0.24(12)(9) & 0.131(9)(9) & 0.1(2)(1) &
    1.(1)(2) \
    & 0.56(10)(3) & 1.(1)(1)\\

    40 & 0.451(11)(1) & 0.242(12)(3) & 0.133(9)(8) & 0.1(2)(1) &
    1.(1)(2) \
    & 0.54(10)(3) & 1.(1)(1)\\

    41 & 0.458(11)(5) & 0.26(1)(1) & 0.151(9)(11) & 0.4(2)(3) &
    0.9(1)(4) \
    & 0.3(1)(2) & 0.6(1)(1)\\

    42 & 0.458(9)(8) & 0.26(1)(2) & 0.152(9)(14) & 0.3(2)(4) & \
    0.88(9)(54) & 0.31(9)(26) & 0.6(1)(2)\\

    43 & 0.46(9)(8) & 0.26(1)(2) & 0.153(9)(10) & 0.3(2)(4) &
    0.8(1)(6) & \
    0.3(9)(19) & 0.6(1)(1)\\

    44 & 0.4526(87)(4) & 0.253(10)(8) & 0.144(6)(1) & 0.1(2)(1) & \
    0.9(1)(5) & 0.39(6)(6) & 0.76(7)(2)\\

    45 & 0.453(9)(1) & 0.254(11)(8) & 0.145(6)(5) & 0.1(2)(2) &
    0.9(1)(5) \
    & 0.39(7)(12) & 0.75(7)(8)\\

    46 & 0.453(9)(3) & 0.25(1)(1) & 0.145(7)(1) & 0.1(2)(2) &
    0.9(1)(5) & \
    0.38(7)(10) & 0.75(8)(6)\\

    47 & 0.453(9)(1) & 0.254(10)(7) & 0.145(7)(4) & 0.1(2)(1) &
    0.9(1)(4) \
    & 0.38(6)(14) & 0.75(7)(10)\\

    48 & 0.454(8)(3) & 0.25(1)(1) & 0.155(7)(13) & 0.085(20)(5) & \
    0.9(1)(5) & 0.35(8)(13) & 0.6(9)(22)\\

    49 & 0.453(9)(6) & 0.25(1)(1) & 0.155(7)(14) & 0.1(19)(4) &
    0.9(1)(4) \
    & 0.35(8)(14) & 0.61(9)(22)\\

    50 & 0.453(9)(4) & 0.25(1)(1) & 0.156(8)(10) & 0.11(21)(0) & \
    0.9(1)(5) & 0.34(8)(11) & 0.6(1)(2)\\

    51 & 0.462(6)(14) & 0.27(1)(3) & 0.173(5)(19) & 0.09(1)(3) & \
    0.6(1)(8) & 0.14(3)(23) & 0.35(4)(34)\\

    52 & 0.462(7)(15) & 0.27(1)(3) & 0.172(5)(20) & 0.1(1)(3) &
    0.7(1)(8) \
    & 0.14(4)(25) & 0.35(4)(37)\\

    53 & 0.462(6)(16) & 0.273(9)(31) & 0.172(4)(18) & 0.1(1)(3) & \
    0.7(1)(7) & 0.13(2)(23) & 0.35(3)(33)\\

    54 & 0.458(5)(11) & 0.269(9)(29) & 0.168(4)(14) & 0.07(1)(2) & \
    0.7(1)(7) & 0.18(2)(19) & 0.38(3)(30)\\

    55 & 0.459(7)(13) & 0.27(1)(2) & 0.171(6)(15) & 0.04(1)(1) & \
    0.6(2)(8) & 0.18(3)(18) & 0.37(4)(29)\\

    56 & 0.464(5)(16) & 0.28(1)(3) & 0.175(5)(19) & 0.007(1)(2) & \
    0.5(2)(6) & 0.16(1)(22) & 0.33(3)(31)\\

    57 & 0.464(6)(12) & 0.28(1)(2) & 0.175(5)(19) & 0.0029(10)(1) & \
    0.5(2)(5) & 0.16(1)(13) & 0.33(3)(35)\\

    58 & 0.466(8)(10) & 0.28(1)(1) & 0.175(6)(9) & 0.0625(67)(9) & \
    0.5(2)(5) & 0.15(2)(11) & 0.32(4)(19)\\

    59 & 0.465(8)(13) & 0.27(1)(2) & 0.174(6)(15) & 0.049(6)(14) & \
    0.5(2)(5) & 0.15(2)(14) & 0.32(5)(26)\\

    60 & 0.465(8)(19) & 0.279(9)(27) & 0.174(6)(18) & 0.049(6)(24) & \
    0.5(2)(5) & 0.15(2)(18) & 0.33(4)(25)\\

    61 & 0.465(8)(19) & 0.279(9)(27) & 0.174(6)(18) & 0.049(6)(24) & \
    0.5(2)(5) & 0.15(2)(18) & 0.33(4)(25)\\

    62 & 0.46(1)(1) & 0.28(1)(3) & 0.176(7)(17) & 0.047(13)(1) & \
    0.5(1)(7) & 0.14(4)(16) & 0.31(4)(36)\\

    63 & 0.468(9)(22) & 0.28(1)(3) & 0.177(7)(20) & 0.06(1)(4) & \
    0.4(1)(7) & 0.13(4)(22) & 0.3(4)(33)\\

    64 & 0.468(9)(14) & 0.28(1)(3) & 0.177(7)(16) & 0.06(1)(1) & \
    0.4(1)(6) & 0.13(4)(15) & 0.3(4)(21)\\

    65 & 0.468(9)(12) & 0.28(1)(3) & 0.177(7)(18) & 0.065(13)(6) & \
    0.4(1)(5) & 0.13(4)(13) & 0.3(4)(26)\\

    66 & 0.468(8)(16) & 0.28(1)(3) & 0.177(6)(17) & 0.06(1)(2) & \
    0.4(1)(6) & 0.13(3)(17) & 0.29(3)(24)\\

    67 & 0.468(8)(16) & 0.28(1)(3) & 0.177(6)(17) & 0.06(1)(2) & \
    0.4(1)(6) & 0.13(3)(17) & 0.29(3)(24)\\

    68 & 0.468(7)(14) & 0.28(1)(3) & 0.177(6)(15) & 0.06(1)(1) & \
    0.4(1)(6) & 0.13(3)(12) & 0.29(3)(24)\\

    69 & 0.467(5)(15) & 0.28(1)(3) & 0.177(6)(15) & 0.08(1)(2) & \
    0.46(9)(63) & 0.12(3)(14) & 0.29(3)(20)\\

    70 & 0.467(6)(13) & 0.28(1)(3) & 0.177(6)(16) & 0.085(9)(13) & \
    0.45(8)(68) & 0.12(3)(12) & 0.28(2)(20)\\

    71 & 0.467(6)(19) & 0.28(1)(3) & 0.177(6)(21) & 0.087(8)(26) & \
    0.45(8)(64) & 0.12(3)(20) & 0.28(2)(28)\\

    72 & 0.468(6)(14) & 0.28(1)(3) & 0.177(6)(15) & 0.1(1)(1) &
    0.4(1)(6) \
    & 0.11(4)(14) & 0.27(3)(19)\\

    73 & 0.467(8)(13) & 0.28(1)(3) & 0.177(6)(16) & 0.1(1)(1) & \
    0.43(9)(61) & 0.11(3)(16) & 0.27(3)(21)\\

    74 & 0.467(8)(18) & 0.28(1)(3) & 0.177(6)(17) & 0.09(1)(3) & \
    0.41(9)(67) & 0.11(3)(17) & 0.27(3)(23)\\

    75 & 0.466(7)(19) & 0.28(1)(3) & 0.177(6)(16) & 0.08(1)(3) & \
    0.43(9)(62) & 0.12(4)(18) & 0.27(3)(24)\\

    76 & 0.464(7)(17) & 0.28(1)(3) & 0.175(7)(18) & 0.04(1)(2) & \
    0.4(1)(6) & 0.13(5)(16) & 0.29(4)(25)\\

    77 & 0.46(1)(1) & 0.28(1)(2) & 0.175(7)(11) & 0.04(1)(4) &
    0.4(1)(5) \
    & 0.11(3)(15) & 0.26(2)(28)\\

    78 & 0.46(1)(1) & 0.27(2)(3) & 0.175(9)(14) & 0.01(1)(2) &
    0.4(1)(7) \
    & 0.12(4)(13) & 0.26(3)(25)\\

    79 & 0.46(1)(1) & 0.27(2)(2) & 0.175(9)(12) & 0.01(1)(3) &
    0.4(1)(6) \
    & 0.12(4)(19) & 0.26(3)(24)\\

    80 & 0.462(14)(3) & 0.27(2)(1) & 0.175(9)(11) & 0.017(19)(7) & \
    0.4(1)(5) & 0.12(4)(11) & 0.27(3)(21)\\

    81 & 0.46(1)(1) & 0.27(1)(3) & 0.175(8)(12) & 0.01(1)(2) &
    0.4(1)(6) \
    & 0.11(4)(13) & 0.26(3)(18)\\

    82 & 0.46(1)(1) & 0.27(1)(3) & 0.174(8)(15) & 0.03(2)(2) &
    0.4(1)(6) \
    & 0.11(4)(16) & 0.26(3)(21)\\

    83 & 0.46(1)(1) & 0.27(1)(3) & 0.174(8)(14) & 0.03(2)(2) &
    0.4(1)(6) \
    & 0.11(4)(18) & 0.27(3)(22)\\

    84 & 0.462(13)(7) & 0.27(1)(1) & 0.174(8)(12) & 0.05(1)(2) & \
    0.4(1)(5) & 0.11(4)(22) & 0.26(3)(27)\\

    85 & 0.46(10)(9) & 0.27(1)(3) & 0.174(6)(16) & 0.019(16)(3) & \
    0.51(7)(65) & 0.12(4)(17) & 0.28(2)(24)\\

    86 & 0.46(10)(9) & 0.27(1)(3) & 0.173(7)(12) & 0.002(1)(1) & \
    0.51(9)(66) & 0.12(4)(12) & 0.28(3)(17)\\

    87 & 0.46(1)(1) & 0.27(1)(1) & 0.173(7)(13) & 0.002(1)(2) & \
    0.51(9)(42) & 0.12(4)(16) & 0.28(3)(20)\\

    88 & 0.46(10)(5) & 0.27(1)(2) & 0.173(7)(12) & 0.002(1)(1) & \
    0.51(9)(59) & 0.12(4)(16) & 0.28(3)(26)\\

    89 & 0.46(9)(6) & 0.27(1)(2) & 0.173(6)(9) & 0.0022(16)(9) & \
    0.51(9)(44) & 0.12(4)(9) & 0.28(2)(13)\\

    90 & 0.461(9)(10) & 0.27(1)(2) & 0.174(6)(21) & 0.02(1)(1) & \
    0.49(9)(77) & 0.12(4)(28) & 0.27(3)(34)\\[2ex]

  \end{longtable}
\end{center}

\section{Example Multi-Meson Correlation Function}
\begin{eqnarray}\label{multiMesonCorrFxnEx}
  C_{4\pi,3K}(t) &=&\frac{1}{144} \mathrm{tr}[K]^3 \mathrm{tr}[\Pi ]^4-\frac{1}{48} \mathrm{tr}[K] \mathrm{tr}\left[K^2\right] \mathrm{tr}[\Pi ]^4+\frac{1}{72} \mathrm{tr}\left[K^3\right] \mathrm{tr}[\Pi ]^4-\frac{1}{24} \mathrm{tr}[K]^3 \mathrm{tr}[\Pi ]^2 \mathrm{tr}\left[\Pi ^2\right] \nonumber \\ 
  &&+\frac{1}{8} \mathrm{tr}[K] \mathrm{tr}\left[K^2\right] \mathrm{tr}[\Pi]^2 \mathrm{tr}\left[\Pi ^2\right]-\frac{1}{12} \mathrm{tr}\left[K^3\right] \mathrm{tr}[\Pi ]^2 \mathrm{tr}\left[\Pi ^2\right]+\frac{1}{48} \mathrm{tr}[K]^3
  \mathrm{tr}\left[\Pi ^2\right]^2 \nonumber \\ 
  &&-\frac{1}{16} \mathrm{tr}[K] \mathrm{tr}\left[K^2\right] \mathrm{tr}\left[\Pi ^2\right]^2+\frac{1}{24} \mathrm{tr}\left[K^3\right]
  \mathrm{tr}\left[\Pi ^2\right]^2+\frac{1}{18} \mathrm{tr}[K]^3 \mathrm{tr}[\Pi ] \mathrm{tr}\left[\Pi^3\right] \nonumber \\ 
  &&-\frac{1}{6} \mathrm{tr}[K] \mathrm{tr}\left[K^2\right]
  \mathrm{tr}[\Pi ] \mathrm{tr}\left[\Pi ^3\right]+\frac{1}{9} \mathrm{tr}\left[K^3\right] \mathrm{tr}[\Pi ] \mathrm{tr}\left[\Pi^3\right]-\frac{1}{24} \mathrm{tr}[K]^3
  \mathrm{tr}\left[\Pi ^4\right]\nonumber \\
  &&+\frac{1}{8} \mathrm{tr}[K] \mathrm{tr}\left[K^2\right] \mathrm{tr}\left[\Pi^4\right]-\frac{1}{12} \mathrm{tr}\left[K^3\right]
  \mathrm{tr}\left[\Pi ^4\right]-\frac{1}{12} \mathrm{tr}[K]^2 \mathrm{tr}[\Pi ]^3 \mathrm{tr}[K\Pi ]\nonumber \\
  &&+\frac{1}{12} \mathrm{tr}\left[K^2\right] \mathrm{tr}[\Pi]^3 \mathrm{tr}[K\Pi ]+\frac{1}{4} \mathrm{tr}[K]^2 \mathrm{tr}[\Pi ] \mathrm{tr}\left[\Pi ^2\right] \mathrm{tr}[K\Pi ]-\frac{1}{4} \mathrm{tr}\left[K^2\right]
  \mathrm{tr}[\Pi ] \mathrm{tr}\left[\Pi ^2\right] \mathrm{tr}[K\Pi]\nonumber \\
  &&-\frac{1}{6} \mathrm{tr}[K]^2 \mathrm{tr}\left[\Pi^3\right] \mathrm{tr}[K\Pi ]+\frac{1}{6}
  \mathrm{tr}\left[K^2\right] \mathrm{tr}\left[\Pi ^3\right] \mathrm{tr}[K\Pi ]+\frac{1}{4} \mathrm{tr}[K] \mathrm{tr}[\Pi ]^2 \mathrm{tr}[K\Pi ]^2\nonumber \\
  &&-\frac{1}{4}\mathrm{tr}[K] \mathrm{tr}\left[\Pi ^2\right] \mathrm{tr}[K\Pi]^2-\frac{1}{6} \mathrm{tr}[\Pi ] \mathrm{tr}[K\Pi ]^3+\frac{1}{4} \mathrm{tr}[K]^2 \mathrm{tr}[\Pi]^2 \mathrm{tr}\left[K\Pi ^2\right]\nonumber \\
  &&-\frac{1}{4} \mathrm{tr}\left[K^2\right] \mathrm{tr}[\Pi ]^2 \mathrm{tr}\left[K\Pi ^2\right]-\frac{1}{4} \mathrm{tr}[K]^2
  \mathrm{tr}\left[\Pi ^2\right] \mathrm{tr}\left[K\Pi ^2\right]+\frac{1}{4} \mathrm{tr}\left[K^2\right] \mathrm{tr}\left[\Pi ^2\right] \mathrm{tr}\left[K\Pi^2\right]\nonumber \\
  &&-\mathrm{tr}[K] \mathrm{tr}[\Pi ] \mathrm{tr}[K\Pi ] \mathrm{tr}\left[K\Pi ^2\right]+\frac{1}{2} \mathrm{tr}[K\Pi ]^2 \mathrm{tr}\left[K\Pi ^2\right]+\frac{1}{2}
  \mathrm{tr}[K] \mathrm{tr}\left[K\Pi ^2\right]^2\nonumber \\
  &&-\frac{1}{2} \mathrm{tr}[K]^2 \mathrm{tr}[\Pi ] \mathrm{tr}\left[K\Pi ^3\right]+\frac{1}{2} \mathrm{tr}\left[K^2\right]
  \mathrm{tr}[\Pi ] \mathrm{tr}\left[K\Pi ^3\right]+\mathrm{tr}[K] \mathrm{tr}[K\Pi ] \mathrm{tr}\left[K\Pi ^3\right]\nonumber \\
  &&+\frac{1}{2} \mathrm{tr}[K]^2 \mathrm{tr}\left[K\Pi^4\right]-\frac{1}{2} \mathrm{tr}\left[K^2\right] \mathrm{tr}\left[K\Pi^4\right]+\frac{1}{6} \mathrm{tr}[K] \mathrm{tr}[\Pi ]^3 \mathrm{tr}\left[K^2\Pi \right] \nonumber \\
  &&-\frac{1}{2} \mathrm{tr}[K] \mathrm{tr}[\Pi ] \mathrm{tr}\left[\Pi^2\right] \mathrm{tr}\left[K^2 \Pi \right]+\frac{1}{3} \mathrm{tr}[K] \mathrm{tr}\left[\Pi^3\right] \mathrm{tr}\left[K^2\Pi \right]-\frac{1}{2} \mathrm{tr}[\Pi]^2 \mathrm{tr}[K\Pi ] \mathrm{tr}\left[K^2\Pi \right]\nonumber \\
  &&+\frac{1}{2} \mathrm{tr}\left[\Pi^2\right] \mathrm{tr}[K\Pi ] \mathrm{tr}\left[K^2\Pi \right]+\mathrm{tr}[\Pi ] \mathrm{tr}\left[K\Pi ^2\right] \mathrm{tr}\left[K^2\Pi \right]-\mathrm{tr}\left[K\Pi^3\right] \mathrm{tr}\left[K^2\Pi \right]\nonumber \\
  &&-\frac{3}{4} \mathrm{tr}[K] \mathrm{tr}[\Pi ]^2 \mathrm{tr}\left[K^2\Pi ^2\right]+\frac{3}{4} \mathrm{tr}[K] \mathrm{tr}\left[\Pi^2\right] \mathrm{tr}\left[K^2\Pi ^2\right]+\frac{3}{2} \mathrm{tr}[\Pi] \mathrm{tr}[K\Pi ] \mathrm{tr}\left[K^2\Pi ^2\right]\nonumber \\
  &&-\frac{3}{2} \mathrm{tr}\left[K\Pi^2\right] \mathrm{tr}\left[K^2\Pi ^2\right]+2 \mathrm{tr}[K] \mathrm{tr}[\Pi ] \mathrm{tr}\left[K^2\Pi ^3\right]-2 \mathrm{tr}[K\Pi ] \mathrm{tr}\left[K^2\Pi^3\right]\nonumber \\
  &&-\frac{5}{2} \mathrm{tr}[K] \mathrm{tr}\left[K^2\Pi ^4\right]-\frac{1}{6} \mathrm{tr}[\Pi]^3 \mathrm{tr}\left[K^3\Pi \right]+\frac{1}{2} \mathrm{tr}[\Pi] \mathrm{tr}\left[\Pi ^2\right] \mathrm{tr}\left[K^3\Pi \right]\nonumber \\
  &&-\frac{1}{3} \mathrm{tr}\left[\Pi ^3\right] \mathrm{tr}\left[K^3\Pi \right]+\mathrm{tr}[\Pi]^2 \mathrm{tr}\left[K^3\Pi ^2\right]-\mathrm{tr}\left[\Pi ^2\right] \mathrm{tr}\left[K^3\Pi ^2\right]-\frac{10}{3} \mathrm{tr}[\Pi ] \mathrm{tr}\
  \left[K^3\Pi^3\right]\nonumber \\
  &&+5 \mathrm{tr}\left[K^3\Pi ^4\right]
\end{eqnarray}

\bibliography{masterBibFile}

\end{document}